\documentclass{aa}  

\usepackage{graphicx}
\usepackage{natbib}
\usepackage{txfonts}
\usepackage{hyperref}
\newcommand{\msun}{{\rm M}_\odot}
\newcommand{\lsun}{{\rm L}_\odot}

\newcommand{\deltal}{\Delta \log{(L_{\rm TRGB}}/\lsun)}
\newcommand{\ltip}{L_{\rm TRGB}}
\newcommand{\dtmin}{\Delta t_{\rm min}}
\newcommand{\teff}{T_{\rm eff}}
\newcommand{\feh}{{\rm [Fe/H]}}

\begin{document}

   \title{The brightness of the Red Giant Branch tip}

   \subtitle{Theoretical framework, a set of reference models, and predicted observables}

   \author{A. Serenelli \inst{1}
          \and A. Weiss \inst{2}
          \and S. Cassisi \inst{3}
          \and M. Salaris \inst{4}
          \and A. Pietrinferni \inst{3}
          }

   \institute{Institute of Space Sciences (IEEC-CSIC) Campus
     UAB, Carrer de Can Magrans, s/n, E-08193, Barcelona,
     Spain \email{aldos@ice.csic.es} \and Max-Planck-Institut f\"ur
     Astrophysik, Karl-Schwarzschild-Str.~1, 86748 Garching, Germany
     \and INAF-Osservatorio Astronomico di Teramo, via M. Maggini,
     64100, Teramo, Italy \and Astrophysics Research Institute,
     Liverpool John Moores University, IC2, Liverpool Science Park,
     146 Brownlow Hill, Liverpool, L3 5RF, UK }

   \date{Received ; accepted }

 
   \abstract
 {The brightness of the tip of
 the Red Giant Branch is a useful reference quantity for several
 fields of astrophysics. An accurate theoretical prediction is needed
 for such purposes.}
 {We intend to provide a solid theoretical prediction for it, valid for a
 reference set of standard physical assumptions, and mostly
 independent of numerical details.}
 {We examine the dependence on physical assumptions and numerical
 details, for a wide range of metallicities and masses, and
 based on two different stellar evolution codes. We adjust
 differences between the codes to treat the physics as
 identical as possible. After we have succeeded in reproducing
 the tip brightness between the codes, we present a reference
 set of models based on the most up to date physical inputs, but neglecting 
 microscopic diffusion, and convert theoretical luminosities to observed
 infrared colours suitable for observations of resolved
 populations of stars and include analytic fits to facilitate their use.}
 {We find that consistent use of updated nuclear reactions, including an appropriate
 treatment of the electron screening effects, and careful time-stepping on the upper
 red giant branch are the most important aspects to bring
 initially discrepant theoretical values into agreement. Small,
 but visible differences remain unexplained for very low
 metallicities and mass values at and above $1.2\,M_\odot$, corresponding to ages
 younger than 4~Gyr. The colour
 transformations introduce larger uncertainties than the
 differences between the two stellar evolution codes. }
 {We demonstrate that careful
stellar modeling allows an accurate prediction for the
luminosity of the Red Giant Branch tip. Differences to
empirically determined brightnesses may result either from
insufficient colour transformations or from deficits in the
constitutional physics. We present the best-tested
theoretical reference values to date.}


   \maketitle
%

\section{Introduction}\label{s:intro}

Stars of low mass, which are in the state of shell hydrogen burning
and evolve along the first red giant branch (RGB), are of a comparably
simple structure, which can be understood to large extent by simple
analytical relations. The method of {\em shell source homology} by 
\citet{refsdalweigert:1970} allows to reproduce the strict connection
between the mass of the degenerate, hydrogen-exhausted helium core
($M_\mathrm{c}$) and the stellar luminosity ($L$), as well as with the
temperature of the shell, which is a first-order approximation for the
maximum temperature inside the core \citep[see][for a detailed derivation
  and further analytical relations]{kiweiwei}. These relations are
formally independent of the total stellar mass within a mass range
of  $\approx 0.5 -  2.0~M_\odot$, which is the classical definition of
low-mass stars.

Due to a rather narrowly defined helium-burning ignition temperature
range close to $10^8$~K, there is thus a
very strong connection between core mass, core temperature and 
luminosity. As the (off-centre) ignition of helium burning, the
so-called core helium flash, leads to a drastic change in the 
stellar interior, notably to a complete lifting of degeneracy, the
star quickly leaves the RGB and re-appears, in the phase of quiescent
central helium burning, at lower luminosity and somewhat higher
effective temperature on what is called either the {\em red clump}, or,
for metal-poor populations, the {\em horizontal branch}. This results
in a sharp upper luminosity limit of the RGB, the tip of the Red Giant
Branch (TRGB).

A well-defined luminosity or absolute brightness for an easily
identifiable class of objects, which, additionally, is only weakly dependent
on composition and mass (or equivalently age), is an irresistible gift for astronomers. 
This is reinforced because, in particular, it also turned out that in 
suitable filters the luminosity of the TRGB is
only weakly dependent on composition \citep{dca:1990,lfm:1993}.

There are two major applications of this fact. The first, and most
obvious one is to employ the TRGB-brightness as a standard candle. For
any given population of a sufficient number of stars, the TRGB can be
easily detected in a resolved colour-magnitude-diagram (CMD) out to
distances of several Mpc \citep[e.g.][]{mcm:2002} by using statistical
filters to identify a break in the luminosity function
\citep{smf:1996}. It was shown repeatedly that the TRGB luminosity can
compete in accuracy with other distance indicators such as
Cepheids, surface brightness fluctuation, and the Planetary Nebulae
Luminosity Function. It has become a standard rung in the
extragalactic distance ladder, and has been used to determine
distances of up to 10~Mpc in over 40 cases \citep{jrt:2009}.

The second method to exploit the stability of the TRGB-brightness uses
the inverse approach: for a stellar system with well-known distance, 
the abolute brightness of the TRGB can be measured and compared to the
theoretical predictions. This can identify deficits in the models, but
the real benefit is to put bounds on additional, postulated physics,
which would influence the details of helium ignition. The method is
mostly applied to globular clusters and in connection with the
properties of neutrinos and axions \citep[and
  others]{rw:1992,rw:1995,cfh:1996}. Here, additional cooling channels
beyond the normal plasma neutrino emission require an increased core
mass to achieve ignition temperatures and therefore, via the core mass
-- luminosity relation, a higher TRGB-luminosity. The challenge for
this method is the low number of stars on the upper RGB, which allows
only for a statistical determination of the TRGB-brightness. More
recently, \citet{vcsrrvw:2013b,vcsrrvw:2013a}  and \citet{adszj:2015}
have used the populous  clusters M5 and $\omega$~Cen for this
purpose.

For both purposes, accurate and reliable theoretical predictions about
the TRGB-luminosity and consequently brightness in various photometric 
bands are needed. Stellar models are computed under a variety of options
for physical effects, of input data, and parameters. Examples are the
inclusion or neglect of atomic diffusion, the various sources for
opacities, and the mixing-length parameter for convection. Therefore,
the dependence of the prediction on these choices are crucial as
well. \citet{rw:1992} and \citet{vcsrrvw:2013b} have studied the
impact of theoretical uncertainties in stellar models have on
photometric predictions, but only for one specific choice
of stellar mass and composition in each work. We will review this in
Sect.~\ref{s:review} along with earlier results of ourselves and new
calculations when appropriate.  
A further summary of the general dependency on mass and composition can be found 
in \citet{scbook:2006}, and a presentation of the stability of the TRGB-brightness 
in the $I$-band in \citet{csbook:2013}.

In Sect.~\ref{s:modcomp} we will compare the results of two
different stellar evolution codes, identify the reasons for the
original discrepancies and resolve them to a large degree. With a
satisfying agreement being achieved, we present a representative set
of models for well-defined, generic physics assumptions in
Sect.~\ref{s:standard}. 
This will be followed by a discussion of the theoretical calculations in 
the context of using the TRGB as distance indicator, studying the additional effect 
of the choice of bolometric corrections, and including comparisons of our results 
with empirical constraints in Sect.~\ref{s:distance}.
We finally discuss in Sect.~\ref{s:discussion} the
remaining uncertainties and the overal reliability of our reference
predictions in the conclusions.

\section{The dependence of the TRGB brightness on input physics} \label{s:review}

Due to the existence of a well-established He core mass - luminosity
relation (see. e.g., \citealt{csbook:2013}, and references therein)
for RGB models, for any given initial chemical
composition the TRGB bolometric luminosity ($L_{\rm TRGB}$) is 
related to the mass size of the He core at the
He-ignition (${\rm M_c^{He}}$):
 
\[{\rm  {\partial{\log{(L_{TRGB}/L_\odot)}}\over\partial{\log{(M_c^{He}/M_\odot)}}}\sim6}.\]

This strong dependence implies that even a small change in the mass of the He core at
the TRGB has a sizable effect on the luminosity of the models.

Theoretical ${\rm M_c^{He}}$ values depend 
on both input physics and numerical assumptions in the 
model computations, as shown in the following.
We start discussing the effect on  ${\rm M_c^{He}}$ (and $L_{\rm TRGB}$) 
of varying, one at a time, the most relevant physics inputs, by reviewing 
results available in the
literature, covering mostly the cases of low mass stars ($M<1\,\msun$)
and low metallicities (typically $Z < 0.006$).
We refer mostly to \citet{cfh:1996,ccdw:98, scbook:2006, csbook:2013,
  vcsrrvw:2013a, valle:13}. 
If no reference is given, the results stem from unpublished tests
made previously by the current authors.

In this work we have
extended both mass and metallicity ranges, including 
models up to 1.4~$\msun$ and Z= 0.04 to include all possible cases for TRGBs 
as young as 4~Gyr. Results for this extended set of models will be discussed in
the following sections.

\subsection{Nuclear reaction rates}

Nuclear reaction rates are a key ingredient in the calculation of
stellar models, and much effort has been devoted to improve the
measurements of rates at energies as close as possible to the Gamow
peak, i.e., the energies at which nuclear reactions occur in
stars. Thanks to these studies, the reaction rates involved in the
\emph{p-p chain} are now established with small uncertainties. The same
is true for many reactions of 
the \emph{CNO-cycle}, but for the ${\rm ^{14}N(p, \gamma)^{15}O}$
rate, that could be still affected by some lingering
uncertainty. This reaction is the slowest one in 
the whole cycle and therefore controls the \emph{CNO-cycle}
efficiency. It is important in low-mass stars both near the end
of core H-burning \citep{imbriani:04} and during the
RGB evolution, because H-burning in the advancing shell is 
controlled by the \emph{CNO-cycle}.

In the past, the rate for the ${\rm ^{14}N(p, \gamma)^{15}O}$ reaction
was affected by a significant uncertainty. The situation has 
improved thanks to the LUNA experiment \citep{formicola:04} that 
provided a more accurate determination of the cross section at
relevant stellar energies. The updated rate is about a
factor of two lower than older estimates, such as the NACRE one
\citep{angulo:99} still widely used in the literature.

The impact of the LUNA rate on the He core mass and brightness at
the RGB tip of low-mass stars has been
investigated by \citet{weiss:05} and \citet{pcs:10}. \citet{pcs:10}
found that the LUNA reaction rate leads to a larger value for ${\rm
  M_c^{He}}$, of the order of ${\rm \approx0.002 - 0.003M_\odot}$,
with respect to the results obtained by using the NACRE rate. Despite
the larger He core mass of the models based on the LUNA rate, their
TRGB brightness is lower by about ${\Delta\log(L_{\rm
    TRGB}/\lsun)=0.02}$. This is due to the fact that the
lower \emph{CNO-cycle} efficiency in the H-burning shell of the LUNA
models compensates for the luminosity increase expected from the
larger core mass, as  discussed by \citet{weiss:05}. 
The impact of this brightness change on the absolute I-Cousins 
magnitude of the TRGB is about 0.05~mag for a metallicity
below ${\rm [Fe/H]\approx -1.0}$, and even smaller at larger
metallicities\footnote{If the metal mixture is fixed in the models, and  ${\rm [\alpha/Fe]=0}$, 
then ${\rm [Fe/H]=[M/H]=log(Z/X)-log(Z/X)_{\odot}}$, where the solar value of the
  ratio Z/X is taken from the solar metal
  distribution adopted in the model calculations.}.
In terms of simple analytical models, the effect can be understood
as follows: for given luminosity $L$, the lower \emph{CNO-cycle}
efficiency requires a higher burning temperature in the shell. As this
temperature sets the scale of the core temperature, the core will
reach the He-ignition temperature at a lower luminosity (mass).

More recently this issue has again been investigated by
\citet{vcsrrvw:2013a} who checked the impact on $M_I^{\rm TRGB}$ due
to a variation of $\pm15$\% of the ${\rm ^{14}N(p, \gamma)^{15}O}$
reaction rate around the LUNA value. For their case study, an
$M=0.82\,\msun$ and $Z=0.00136$ model, they found this uncertainty to
affect $M_I^{\rm TRGB}$ at the level of
$\sim\mp0.009$~mag, a weaker but consistent dependence, also in good
agreement with the M=0.9~$\msun$, Z=0.006 model studied by
\citet{valle:13}.

Since the TRGB marks the He ignition in the stellar core, another
important nuclear process is the triple alpha reaction ${\rm 3\alpha
  \rightarrow{^{12}C+\gamma}}$. The presence of resonances
in the low energy range complicates an accurate estimate
of this reaction rate. The most recent measurement has been obtained by
\citet{fynbo:05}: they found a significant, temperature-dependent
deviation from the older NACRE
value, up to -20\% in the temperature range
$70-100\times10^6\,\hbox{K}$ relevant to He-ignition in RGB
cores. However, the difference decreases towards the upper temperature value,
which is also the canonical He-ignition temperature \citep[see Fig.~1
  of][]{weiss:05}. 
As a consequence, the impact of this new determination on the properties
of TRGB models has turned out to be very small, as verified by
\citet{weiss:05} for low masses and low metallicities. The He core
mass shows a very small increases, between 0.002 and ${\rm
  0.003M_\odot}$ (the exact value depending on the chemical
composition and initial mass), and TRGB luminosity 
increases by about $\Delta\log(L_{\rm TRGB}/\lsun)=0.009$, i.e., by
about 2\%, that translates to a ${\rm \Delta{M_I^{TRGB}}\approx
  0.02}$~mag. This is consistent with the result by
\citet{vcsrrvw:2013a} who found, for their only test case, that a
maximum uncertainty of $\sim30$\% on the updated reaction rate would
introduce a maximum $M_I^{\rm TRGB}$ variation of about 0.02~mag.
This increase of the TRGB luminosity is consistent with the
expectation that a lower $3\alpha$-rate requires a higher core
temperature, i.e.\ a larger core mass, to initiate the core helium flash.

The bottom line of this analysis is that realistic residual
uncertainties in these two reaction rates probably
do not affect significantly the theoretical predictions of the
TRGB absolute magnitude. 

\subsection{Nuclear reaction screening}\label{sec:screening}

Nuclear reactions inside the stars occur in a plasma where free
electrons tend to distribute around the nuclei, shielding the
Coulomb potential felt by the approaching particles. This implies that
a correction -- the so-called electron screening -- has to be applied 
to properly evaluate the actual reaction rates in stellar
models. Despite its relevance, the correct treatment of
electron screening in a stellar plasma is
still an unsettled issue \citep[see, e.g.,][for some discussion on this
  topic]{vcsrrvw:2013a}.

As a consequence, there is no proper estimate of the accuracy of 
the available electron screening predictions. An extreme test is to 
calculate models neglecting electron screening, to be compared 
with results obtained with the appropriate choice between the various
\lq{levels}\rq\ (\emph{strong, intermediate} and \emph{weak}) of
screening, determined by the available theoretical predictions
\citep{dewitt:1973,graboske:1973}. The result of this experiment is
that ${\rm M_c^{He}}$ increases by ${\rm \approx0.02M_\odot}$ when 
screening is neglected, corresponding to $\deltal\approx0.12$, 
hence about 0.30~mag in $M_I^{\rm
  TRGB}$.

\subsection{Equation of state}

The equation of state (EOS) is one of the most important ingredients
in stellar model computations, and in the last decade significant
improvements have been achieved in the description of the 
thermal conditions in the interiors and atmospheres of low-mass
stars. Various EOSs
currently available \citep[see][for a compilation]{vcsrrvw:2013a} 
provide similar predictions; hence the
use of different EOS does not affect the TRGB. 
Indeed, \cite{vcsrrvw:2013a} have shown that, when
compared with models based on the most updated EOS (the
FreeEOS\footnote{\url{http://freeeos.sourceforge.net/}},
\citealp{csi:03}), the use of different EOS prescriptions 
introduces a maximum variation of
$+0.02$~mag in $M_I^{\rm TRGB}$.

Despite this result, it is in principle conceivable that for the   
highest density layers of the models, there could still be EOS 
uncertainties related to the treatment of the Coulomb and electron exchange
effects. Accounting for different implementations of such processes in
the EOS computation is quite difficult, but the FreeEOS code
allows to include or neglect these non-ideal effects. 
This has allowed us to estimate what is most probably an upper limit for 
the uncertainty associated to the EOS in the regime of RGB He cores\footnote{A
  similar approach has been followed by 
  \cite{csi:03} to estimate the effect of EOS uncertainties
  on the theoretical predictions for the $R-$parameter.}. 

Calculations of low-mass star models  
neglecting the electron exchange effects show an
increase of the He core mass  at the TRGB by ${\rm
  \Delta{M_c^{He}}=9\times10^{-4}M_\odot}$, and a decrease of the
surface luminosity by $\Delta\log(L_{\rm TRGB}/\lsun)\approx0.003$. 
The decrease of the luminosity despite 
the (quite small) increase of the He core mass is due to
the change of the inner temperature stratification which affects
(decreases) the efficiency of the shell H-burning. Switching
off the effects associated to Coulomb interactions in the EOS
computations causes an increase of the He core mass by ${\rm
  \Delta{M_c^{He}}=4\times10^{-3}M_\odot}$, while as in the previous
case the luminosity at the He ignition decreases by
$\Delta\log(L_{TRGB}/\lsun)\approx0.003$.  When both Coulomb and
electron exchange effects are neglected in the EOS, 
the He core mass increases 
about ${\rm 0.0054\, M_\odot}$ and TRGB brightness is reduced by 
$\Delta\log(L_{\rm TRGB}/\lsun)\approx0.005$.

These tests show that very probably 
lingering uncertainties in the EOS have a negligible -- if any
-- impact on predictions for the TRGB brightness.
This is in line with the results by
\citet{vcsrrvw:2013a}.

\subsection{Radiative opacity}

Only the high-temperature
(${\rm T>10^6~K}$) radiative Rosseland mean opacities can have some effect 
on the interior properties of RGB stars \citep{scs:93}, although  
opacities in the low-temperature regime affect the ${\rm
  T_{eff}}$ of the models, hence the TRGB magnitudes through the
indirect effect on the bolometric corrections (see Sect.~\ref{s:distance}).

In the electron degenerate He cores of evolved RGB stars 
the dominant energy transport mechanism 
is electron conduction, and as the models climb up the RGB, the contribution
of the radiative opacity to the total opacity in the core becomes 
negligible in comparison with conductive opacity 
\citep[see, e.g.][and references therein]{scbook:2006}. 
As a consequence, present uncertainties in
high-T radiative opacities should
have a marginal impact on the TRGB properties of low-mass
stars. Indeed, this expectation has been confirmed in the
analysis made by \citet{vcsrrvw:2013a}: they found that by changing
the radiative opacity by $\pm10$\% causes a variation 
on $M_I^{\rm TRGB}$ by just  $\approx\mp3\cdot10^{-4}$.
Currently, the uncertainty in the high-T radiative 
opacities is believed to be of the order of $\sim5$\%, although 
there is some claim \citep[see, e.g.,][for a detailed discussion on
this issue]{catelan:13} that this figure could reach a
value around 20-30\% in some specific temperature ranges.

\subsection{Conductive opacity} \label{sec:condopac}

Electron conduction regulates the thermal state of the degenerate He core, 
and a
reliable estimate of the conductive opacities (${\rm \kappa_{cond}}$)
is fundamental for deriving the correct value of ${\rm M_c^{He}}$. As
a rule of thumb, higher conductive opacities cause a less efficient
cooling of the He-core and an earlier He-ignition, hence at a lower
core mass and fainter TRGB brightness.  The most recent ${\rm
  \kappa_{cond}}$ update has been provided by \cite{cppcs:07}: this
opacity set fully covers the thermal conditions characteristic of
electron degenerate cores in low-mass, metal-poor stars, accounts for
arbitrary chemical mixtures, and takes into account the important
contributions of electron-ion scattering and electron-electron
scattering in the regime of partial electron
degeneracy, which indeed  is the relevant condition for the
He cores of low-mass RGB stars. These new ${\rm \kappa_{cond}}$
data result in values of ${\rm M_c^{He}}$ lower than when using the
previous calculations by \cite{p:99}: the difference 
amounts to ${\rm 0.006M_\odot}$, regardless of the two metallicities
tested. This difference in He-core mass at the RGB tip causes a 
TRGB fainter by $\Delta\log(L_{\rm TRGB}/\lsun\approx0.03$
for the models based on the most recent ${\rm \kappa_{cond}}$ values. 

Again, it is difficult to estimate the remaining 
uncertainty in ${\rm \kappa_{cond}}$, because of the
complexity of the various physical processes whose combined effects
determines the conductive transport efficiency. \citet{vcsrrvw:2013a}
and \citet{valle:13} assumed a 10\% and 5\% uncertainty, respectively,
which resulted in variations of 0.016 and 0.007~mag in $M_I^{\rm TRGB}$.


\subsection{Neutrino energy losses}

In the high temperature and density regimes characteristic of low-mass
red giants, neutrino peoduction through plasma-,
photo-, pair-, and bremsstrahlung-processes, is quite efficient.
These energy loss channels play a 
crucial role in determining the value of the TRGB He core mass and 
brightness. In the dense and not extremely hot He core of
RGB stars, plasma neutrino emission is the dominant process.

The most recent calculation of neutrino loss rates are still the ones provided 
by \citet{haft:94} and \citet{itoh:96a}. Although there is some claim
by \citet{itoh:96b} that the accuracy of these rates 
should be around 5\%, there
is no clear proof that these predictions are unaffected by
some systematic errors. To estimate the impact
that potential uncertainties in neutrino energy losses can have on
low-mass stellar models, we compared results based on the
most recent evaluations, with calculations performed using 
the older \citet{munakata:85} rates. To this purpose, we rely on
the numerical experiments performed by \citet{ccdw:98}: these authors
found that the use of the most updated neutrino energy losses 
--keeping fixed all other input physics-- causes an increase of
${\rm M_c^{He}}$ of about ${\rm 0.006\,M_\odot}$, and a 
$\deltal \approx 0.03$, corresponding to a variation of $M_I^{\rm TRGB}$ by 
$\approx-0.08$~mag. This variation is about a 
factor of 6 larger than the effect estimated by \citet{vcsrrvw:2013a}
by assuming a change of $\pm5$\% in the neutrino emission rates by
\citet{haft:94}.

We note that present estimates of the plasma neutrino energy losses
are based on the canonical assumption that neutrinos have no magnetic
moment. If neutrinos have direct electromagnetic interactions as
due to the presence of a magnetic moment, the energy loss rates 
would be strongly enhanced, increasing the efficiency of
neutrino cooling for larger magnetic moment. This would lead to an
increase of the He core mass and luminosity at
the TRGB. As mentioned already in the introduction, several investigations
\citep{raffelt:90,cd:93,vcsrrvw:2013a,vcsrrvw:2013b} have been devoted
to use the TRGB brightness in Galactic globular clusters to set  
upper limits for the neutrino magnetic moment
\citep{vcsrrvw:2013a}. Such investigations necessarily rely on the
assumption that the known neutrino emission processes are treated with a 
sufficiently high accuracy.

\subsection{Diffusive processes}\label{s:diffusion}

Atomic diffusion (element transport due to temperature, pressure and
chemical abundance gradients)  
and radiative levitation (element transport due to the 
momentum imparted to ions by the outgoing photons) 
are potentially very important in low-mass stars, as a consequence of their long 
main sequence lifetimes \citep[see, e.g.,][for a detailed discussion on
  this topic and relevant references]{csbook:2013}. As for the 
properties of the models at the TRGB, it is diffusion 
that can be important, as investigated
by \citet{ccdw:98} and by \citet{mrr:10}. If atomic diffusion is fully
efficient during the  
main sequence phase, the TRGB ${\rm M_c^{He}}$ is increased by about
${\rm 0.003 - 0.004M_\odot}$ for ${\rm Z\le0.001}$, this variation
slightly decreasing when increasing the metallicity. The increase of
${\rm M_c^{He}}$ is a consequence of the small 
decrease ($\Delta {\rm Y} \approx -0.01$) of the He-abundance in the envelope during the RGB phase, compared to models calculated without diffusion.
This reduction changes the mean molecular weight, hence the efficiency of shell H-burning, 
delaying the ignition of the helium core (we remind the reader that ${\rm L_H\propto\mu^7}$; \citealp{kiweiwei}).

Due to the larger He core mass, the TRGB brightness increases 
by $\deltal \sim 0.003$ at Z=0.0002, and $\sim0.01$ at Z=0.002; these variations 
translate to a change $\Delta M_I^{\rm TRGB}$ from 
$-0.01$~up to~$-0.03$~mag \citep[see also][and references
  therein]{vcsrrvw:2013a}. 
  
Before closing this section, we note that the atomic diffusion
velocities are generally thought to be accurate at the level of about
15-20\% \citep{thoul:1994}. However, it is difficult to account for these
uncertainties in stellar model computations because this uncertainty
might not be just a systematic error affecting all  diffusion velocities of
the various chemical elements (but see the analysis performed by
\citealt{valle:13}). Note also that the changes in TRGB 
properties estimated above are likely to be upper
limits because they are based on fully efficient atomic diffusion. 
Additional effects such as radiative levitation or extra-mixing are 
likely to diminish the net effect of atomic diffusion.

\section{Models and comparisons}\label{s:modcomp}

The dependence of $L_{\rm TRGB}$ on the input physics entering
stellar model calculations has been discussed in the previous
section. But, in order to have a better assessment of the robustness
of theoretical predictions for $L_{\rm TRGB}$, it is important to
consider the consistency of results obtained with different
evolutionary codes. This is discussed in this section, based on
results obtained with the \texttt{BaSTI} code
\citep{pietrinferni:2004} and with \texttt{GARSTEC}
\citep{weiss:2008}, in three different steps. Initially, we make a
direct comparison of results obtained from calculations that were
already available. Next, we define a concordance set of 
state-of-the-art input physics and compare stellar models newly
computed based on this physics. Finally, we consider the influence of
numerics that can differ among numerical codes and 
impact the predicted $L_{\rm TRGB}$.

\subsection{Initial comparison of two stellar code results}\label{s:initcomp} 

The initial comparison is based on \texttt{BaSTI} results obtained
using the physics described in \citet{pietrinferni:2004}, and
available at the \texttt{BaSTI}
repository\footnote{http://www.oa-teramo.inaf.it/BASTI/}, and on
\texttt{GARSTEC} models computed using the physics described in
\citet{weiss:2008}.  The most notable differences between the physics
included in these models are: the choice of the ${\rm
  ^{14}N(p,\gamma)^{15}O}$ rate (NACRE for \texttt{BaSTI} and LUNA for
\texttt{GARSTEC}), electron screening (\citealp{graboske:1973} for
\texttt{BaSTI} and weak screening \citep{salpeter:54} for
\texttt{GARSTEC}), and conductive opacity (\citealp{p:99} in
\texttt{BaSTI} and \citealp{cppcs:07} in \texttt{GARSTEC}).

\begin{figure}
   \centering
   \includegraphics[width=9cm]{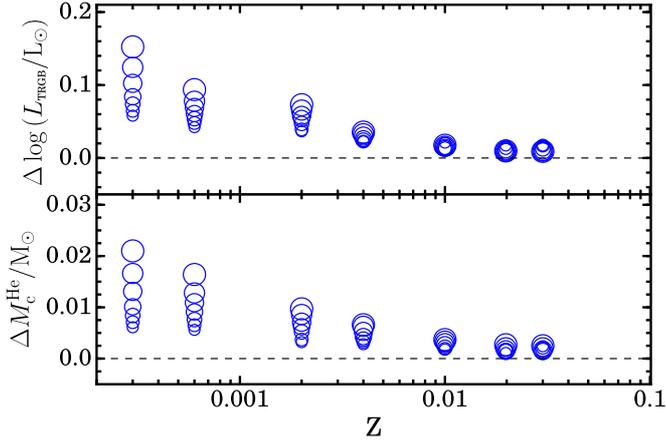}
      \caption{Differences in the TRGB luminosity (top panel) and core
        mass at helium ignition(bottom panel) between two stellar
        codes as a function of metallicity. Symbol sizes denote the
        stellar mass, from 0.8~$\msun$ to 1.4~$\msun$
        (smallest and largest symbols respectively). Differences are
        in the sense \texttt{BaSTI}~$-$~\texttt{GARSTEC}. 
        \label{fig:interinitial}}
   \end{figure}

In Fig.\,\ref{fig:interinitial} the differences in $\deltal$ and
$\rm{M_c^{He}}$, in the sense \texttt{BaSTI} $-$ \texttt{GARSTEC} are
shown as a function of metallicity and for a range of masses between
0.8 and 1.4~$\msun$. At metallicities $Z > 0.01$ there is good
agreement across the whole mass range. This must be considered with
some caution, because the relevant physics is different in both sets of
calculations. As shown later on, this agreement results from the 
compensation of differences between physics and numerics.  At lower
metallicity, differences increase substantially. Moreover, not only
the zero point agreement degrades to 0.05~dex at $Z=0.0003$, but also
a large spread in stellar mass appears. These results show that lower
metallicities and also larger masses imply higher sensitivity of the 
results on the input physics. For reference, in well studied cases
such as the globular cluster M5 \citep{vcsrrvw:2013a}, the observational precision with
which $L_{\rm TRGB}$ can be determined is about 0.04~dex.

\subsection{Two stellar codes, same physics}\label{s:physics}

Differences found in the previous section have two origins:
different physical inputs and numerics in the stellar
evolution codes. The next step is then to isolate each of these
components. To this aim, we have defined a set of concordance and at
the same time state-of-the-art physics 
and performed new ad-hoc calculations with both \texttt{BaSTI} and
\texttt{GARSTEC}.

In detail, we use: low temperature opacities from \citet{ferguson:2005},
atomic radiative opacities from OPAL \citep{opalopac:1996}, conductive
opacities from \citet{cppcs:07}, the FreeEOS equation of state
\citep{csi:03}, neutrino energy losses from \citet{haft:94} for plasma and from
\citet{munakata:85} updated to \citet{itoh:96a} for other neutrino
processes, weak and intermediate nuclear screening following
\citet{dewitt:1973} and \citet{graboske:1973}. Nuclear reaction rates
are not strictly the same. \texttt{GARSTEC} models use the Solar
Fusion II (SFII) rates \citep{adelb} for all H-burning reactions and
NACRE for the 3-$\alpha$ reaction, the only relevant He-burning
reaction in triggering He-core ignition. The basic set of H-burning
reaction rates in \texttt{BaSTI} are from NACRE \citep{nacre} with the
rate of ${\rm ^{14}N(p,\gamma)^{15}O}$ updated to the recommended
value from LUNA. The latter is different from the SFII value by only
about 2\%, a minimal difference. Other rates linked to H-burning have
a very minor influence on the luminosity at which helium is
ignited. Therefore, despite the differences in the sources of the
adopted nuclear reaction rates, in practice both codes use very
similar values for the reactions relevant to this work.
The set of model compositions we use follow those in 
\citet{pietrinferni:2004}, with $\Delta {\rm Y}/\Delta {\rm Z} \sim 1.4$, Y=0.245 at Z=0
and the metal mixture from \citet[hereafter GN93]{gn93}. Convection is modeled using the 
mixing length theory with solar calibrated parameters $\alpha_{\rm MLT}=1.83$ for 
\texttt{GARSTEC} and 2.02 for \texttt{BASTI}. The different calibration values 
are mainly due to the use of an Eddington atmosphere in \texttt{GARSTEC} and Krishna-Swamy in \texttt{BASTI}.

The impact of microscopic diffusion in the TRGB brightness is small. Results discussed in 
Sect.~\ref{s:diffusion} show this even in the assumption that microscopic diffusion is 
fully operational, without considering possible processes such as radiative levitation, 
weak stellar wind during the main sequence phase or extra-mixing below the convective envelope all of
which would diminish the effect of microscopic diffusion even further. While standard solar models 
routinely include microscopic diffusion (e.g. \citealt{vinyoles:2017}), it is also true that helioseismic 
constraints favor a smaller ($\sim$ 10-20\%) effective rate than predicted by fully efficient 
diffusion models \citet{delahaye:2006,villante:2014}. The indication that diffusion is too efficient is 
even more compelling in metal-poor stars and has led to the introduction ad-hoc reciped for 
extra-mixing \citep{richard:2002,vandenberg:2012}. Therefore, after taking into account all these 
considerations, 
we choose not to include microscopic diffusion in our reference models that would have, 
nevertheless a very small impact on the results relevant to this work.

The models computed cover masses in the range 0.8-1.4~$\msun$ and
metal mass fractions $Z$ between $10^{-4}$ and 0.04. This combination
of parameters encompass all cases of interest in which the age of the
TRGB is 4~Gyr or older. In fact, for the more massive and metal poor
models the TRGB is reached in about 2~Gyr.  Results for $\deltal$ are
shown in Fig.\,\ref{fig:intercomp1} as a function of metallicity, and
a clear trend is present. However, there is an overall shift with
respect to the initial comparison shown in
Fig.\,\ref{fig:interinitial}. For $M \leq 1.2~{M_\odot}$, the
agreement is better for the lowest metallicities, for which $\ltip$
from the two codes agree to better than 0.02~dex, and is independent
of stellar mass. This difference
increases up to 0.07~dex at solar and higher metallicities.  We note
that the models of both codes changed according to the expectations
based on the dependencies discussed in Sect.~\ref{s:review}, when
changing the physical inputs to match the concordance physics.
However, for masses above  $M \leq 1.2~{M_\odot}$ 
there is an additional effect, which leads to a spread in $\Delta
\ltip$, especially at $Z < 0.001$. We will come back to this below.

These differences, particularly at high metallicities, are larger than
the combined uncertainty of the individual sources discussed in
Sect.\,\ref{s:review} as indicated in the plot, except for the extreme
case of neglecting electron screening in nuclear reactions
(Sect.\,\ref{sec:screening}). At high metallicities, they are also
much larger than the level of observational uncertainties with which
the brightness of the TRGB can be determined.

\subsection{Numerical stability}\label{s:stability}

The final question we want to address is the impact that  numerics
involved in computation of stellar evolution models have on
predictions related to the TRGB luminosity. Even for well defined
physics inputs, implementation in stellar evolution codes involves,
more often than not, different numerical approaches. From
interpolation of tables (equation of state, opacities) to the
integration of the stellar structure equations (e.g. spatial and
temporal resolution, convergence criteria), numerics have an impact on
the models that stellar evolution codes provide.

We have checked the robustness of the $L_{\rm TRGB}$ predictions against 
changes in the integration time step used
in stellar evolution calculations. This is controlled by
limiting the fractional changes that several physical quantities can
experience in one time step, e.g. temperature and pressure at a given
mass coordinate, surface luminosity, etc. Along the RGB, in
particular, as the stellar luminosity increases, evolution speeds up
and integration time steps become progressively shorter, prompting the
need to compute a larger number of models until the TRGB is
reached. To avoid the calculation of too many stellar models, in
\texttt{GARSTEC} it is possible to limit the smallest allowed time
step, $\dtmin$.

   \begin{figure}
   \centering
   \includegraphics[width=9cm]{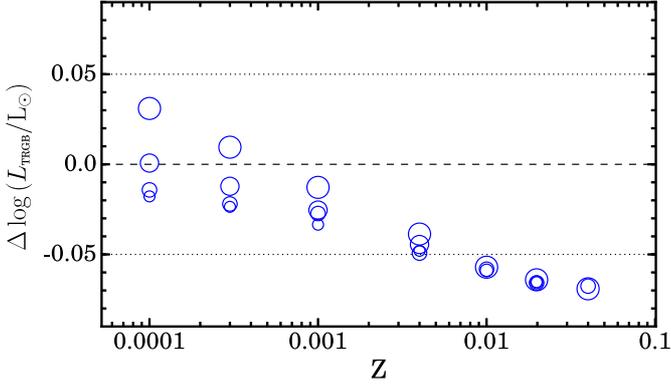}
      \caption{Differences in TRGB luminosity as in top panel of
        Fig.\,\ref{fig:interinitial}, but for models computed using
        the concordance physics. \texttt{GARSTEC} models have been
        computed with $\dtmin=3\,\hbox{kyr}$. Dotted lines indicate
        the estimated 1$\sigma$ theoretical uncertainty excluding
        electron screening (Sect.~\ref{s:review}). 
         \label{fig:intercomp1}}
   \end{figure}

\begin{figure}
   \centering
   \includegraphics[width=9cm]{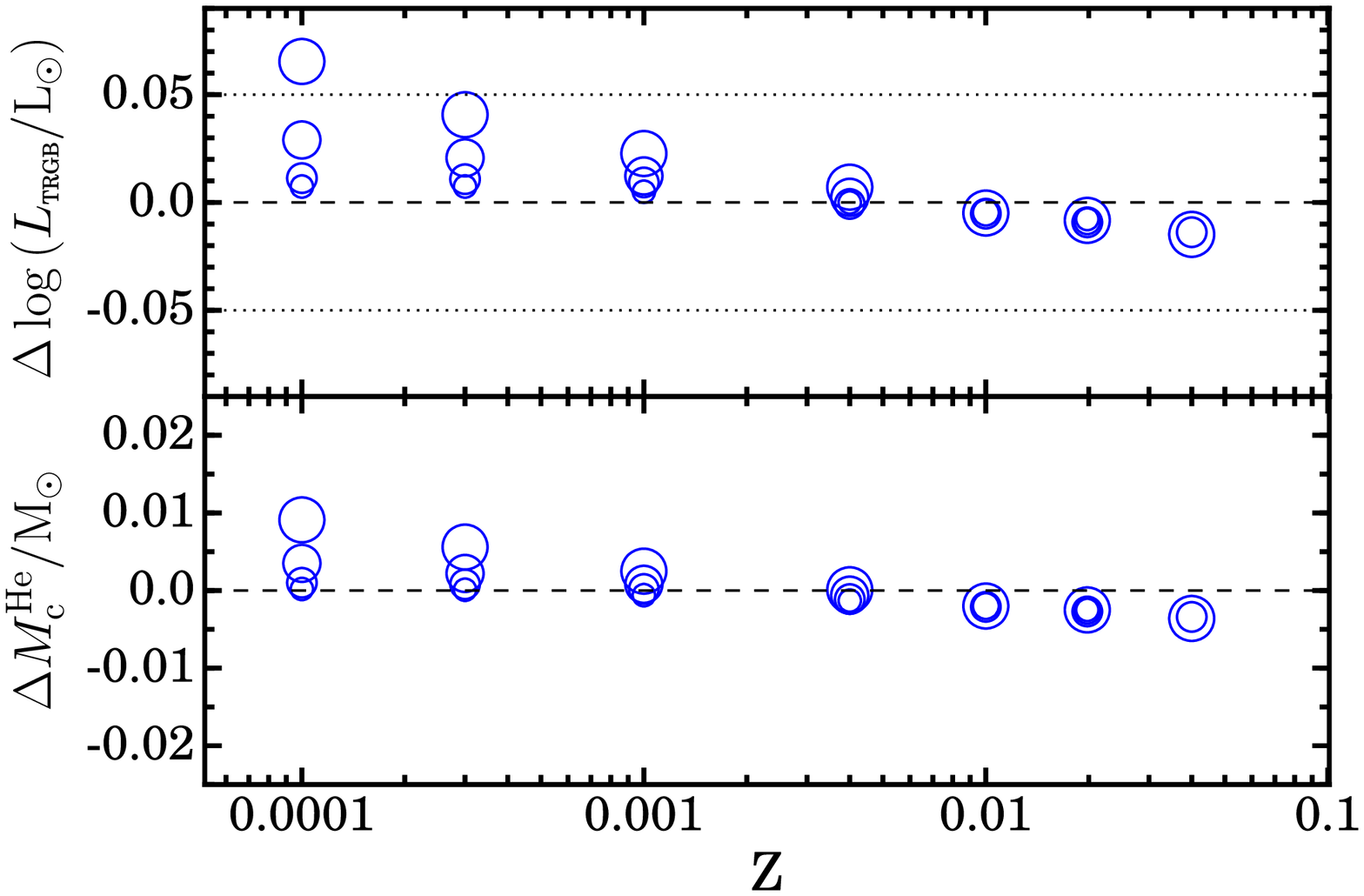}
      \caption{Top panel: same as Fig.\,\ref{fig:intercomp1} but for
        $\dtmin=0.3\,\hbox{kyr}$ in \texttt{GARSTEC}. Bottom panel:
        He-core mass difference.}
         \label{fig:intercomp2}
   \end{figure}

Results in Fig.\,\ref{fig:intercomp1} correspond to calculations where
$\dtmin=3\,\hbox{kyr}$. Along the RGB, in typical \texttt{GARSTEC}
calculations, the constantly decreasing time step reaches this
$\dtmin$ value when $\log{(L_{\rm RGB}/\lsun)} \approx 2.7$. For RGB
calculations, \texttt{BaSTI} does not set a $\dtmin$ value, and only
changes in physical quantities determine the integration time step
that reaches as low as 0.3~kyr close to the TRGB. When a similar minimum timestep
is used in \texttt{GARSTEC}, the comparison between codes is much more
satisfactory. This is shown in Fig.\,\ref{fig:intercomp2}, where the
top panel shows $\deltal$ and the bottom panel differences in $M_{\rm
  c}^{He}$ between both codes. For masses below 1.2~$\msun$ 
differences across all metallicities are now $\deltal < 0.02$, or 
$\Delta M_I^{\rm TRGB} < 0.05$, well within the uncertainties imposed 
by the physics of stellar models. Core masses agree to better than 0.005~$\msun$.

We have carried out additional tests on the dependence of $L_{\rm
  TRGB}$ on the integration time steps used in the codes. The critical
stage is the evolution along the higher RGB, above $\log{(L/\lsun)}
\approx 2.6$ and small integration time steps have to be taken above
this luminosity. In \texttt{GARSTEC}, taking $\dtmin=1\,\hbox{kyr}$ is
enough to yield robust predictions of $L_{\rm TRGB}$; reducing
$\dtmin$ further, even down to $40\,\hbox{yr}$ leads to $\deltal
\approx 0.015$ or, just 0.03 in $M_I^{\rm TRGB}$. With \texttt{BaSTI}
results are similar, $L_{\rm TRGB}$ predictions converge nicely to
within $\deltal \approx 0.01$ for integration time steps that, at the
end of the RGB, are smaller than a few hundred years.

Another important issue concerning numerical stability could be related 
to the choice of an appropriate number of mesh
points in the model computations. This has
been tested by recent numerical experiments performed by one of us
(S.C.): once the integration time step is fixed (in order to disentangle
the effect of mesh resolution from that of time resolution) an
increase of a factor of 2 in the number of spatial mesh points causes
an increase of the He core mass at 
the RGB tip of only $\sim 0.0002M_\odot$, and of the TRGB luminosity of
$\Delta\log(L/L_\odot)\approx0.002$~dex;  a further increase of a
factor of 2 in the number of mesh points does not change the model
predictions at the RGB tip. These effects being quite small, we have
not investigated further this issue in present work, and consider the
standard spatial resolution as being sufficiently fine.

Fig.\,\ref{fig:intercomp2} shows that starting at $Z = 0.004$, 
models with masses $ \geq 1.2~\msun$  show increasing differences with 
decreasing metallicities. This is more noticeable for the 1.4$~\msun$ models, 
the most massive ones shown in the figure, for which differences at the 
lowest metallcities exceed 0.05~dex. This trend is the same as in
Fig.\,\ref{fig:intercomp1}, and therefore does not depend on the time
steps in this phase. In order to understand the origin of this discrepancy 
we have carried out some additional tests. We have tested whether different 
size of the convective cores during the main sequence could affect the RGB-tip luminosity 
for these models. But convective cores for these masses and metallicities are rather small and 
bear no impact on the  RGB evolution. We have also considered the 
different implementation of radiative and conductive opacities in \texttt{GARSTEC} and \texttt{BASTI}, 
particularly in the high density regions and different interpolation schemes for 
radiative opacity (e.g. linear in Z, linear in $\log{{\rm Z}}$). Some 
differences in the final opacities are present in both codes and can be traced back to slightly 
different implementations of the conductive opacities. These differences are of a few percent, 
with \texttt{GARSTEC} being systematically lower than \texttt{BASTI}. But, the temperature-density 
profile in the 1.4~$\msun$ model is very similar to that in a 0.8~$\msun$ model, for which the two 
codes agree very well. In fact, as can be inferred from Sect.~\ref{sec:condopac}, 
a 10\% change in conductive opacities leads to a $\deltal\sim 0.006$~dex, much 
lower than the difference we find for the 1.4~$\msun$ model with Z=0.0001. 
Therefore, we conclude that the systematic opacity difference cannot be the
reason of the different level of agreement between models of different masses at low metallicities.
We have not been able to find the reason for this discrepancy and it certainly requires further 
consideration. Fortunately, in the context of this work, it corresponds to low metallicity stars 
that reach the TRGB in less than 3~Gyr, i.e. in a regime that makes observational interest 
rather limited.

\section{Standard set of models} \label{s:standard}

The choice of concordance physics in the previous section is not only
because it is available both in \texttt{GARSTEC} and \texttt{BaSTI},
but also because it represents state-of-the-art inputs 
appropriate for low-mass stellar models, keeping in mind the uncertainties 
mentioned in Sect.\,\ref{s:physics} regarding microscopic diffusion. We
therefore adopt it as the input physics in the standard set of models
we use in the remainder of this paper and that have been computed
with \texttt{GARSTEC}. In terms of numerics, models have been computed
with a $\dtmin=1~\hbox{kyr}$ minimum allowed time step along the RGB, 
the least restrictive timestep that is
enough to reduce numerical uncertainties well below those from
physical ingredients in the models.  The mass and metallicity ranges
covered by the standard set of models are those used for the intercode
comparison, i.e. M between 0.8 and 1.4~$\msun$ and Z between $10^{-4}$
and 0.04, but with a higher density of models, aimed at including all
possible cases the TRGBs that are 4~Gyr or older.

The composition adopted in  our reference models is based on the GN93 solar 
composition and does not include $\alpha$ enhancement. At fixed $Z$, differences in 
$L_{\rm TRGB}$ and $T_{\rm eff}^{\rm TRGB}$ between models with ${\rm [\alpha/Fe]}=0$  
and 0.4 is less than 1\% and 2\% respectively for masses up to 1.5~$\msun$ and up to 
solar metallicities. Moreover, as shown in \citet{cassisi:2004}, color transformations 
in the infrared filters, $I$, and $(V-I)$ are the same for solar-scaled and 
$\alpha$-enhanced compositions at either the same ${\rm [Fe/H]}$ or ${\rm [M/H]}$, 
with differences being at most 0.01~mag at fixed color in VI and JK color magnitude 
diagrams (see next sections).

The choice of $Y$ in our models assumes a primordial mass fraction of 0.245. Current estimates
range from approximately this value up to 0.256 (see \citealt{cyburt:2016} and references therein). 
A variation of $\Delta Y = 0.01$ has a very small impact in the TRGB properties at all metallicities
and masses of interest in our work, with changes smaller than 1\% in $L_{\rm TRGB}$ and 0.5\% in 
$T_{\rm eff}^{\rm TRGB}$. Therefore, such small variations in $Y$ can be safely neglected as relevant 
sources of uncertainties. A similar argument follows for our choice ${\rm \Delta Y/\Delta Z \sim 1.4}$ because varying ${\rm \Delta Y/\Delta Z}$ in the range from 1 to 2 leads to changes in $Y$ of about 0.01 at solar metallicities and, logically, smaller at lower metallcities.

\begin{figure}
  \centering
   \includegraphics[width=9cm]{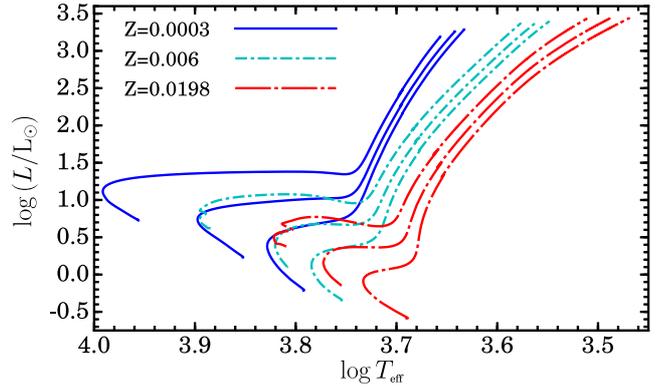}
      \caption{Evolutionary tracks extending from the ZAMS up to the
        TRGB for some of our standard set of models. For each
        metallicity, masses are 0.8, 1, and 1.3~$\msun$.
         \label{fig:tracks}}
   \end{figure}

Fig.\,\ref{fig:tracks} shows several of the evolutionary tracks
corresponding to our standard set of models, for different
metallicities and masses, from the zero age main sequence (ZAMS) up to
the TRGB.

With respect to previous works exploring uncertainties in the
determination of the TRGB luminosity and photometric properties
discussed in Sect.\,\ref{s:review}, we have examined the full mass and
metallicity ranges to assess uncertainties in CMDs for different
choices of photometric bands, complementing the set of standard models
by additional sets of models with varied physics. The photometric
properties of these additional models will be presented in the
next section as well (see Fig.~\ref{vimodels}).

\section{The TRGB as distance indicator} \label{s:distance}

Indications about the possibility of using the TRGB as a distance
indicator can be traced back to \citet{baade} and \citet{sandage}, but
its use for distance determinations started only in the early eighties
\citep[see a summary in][]{lfm:1993}, and the method was formalized by
\citet{lfm:1993}, who proposed the use of an edge-detection algorithm
to identify objectively the TRGB in the CMD of galaxies. A thorough
analysis of the technique and estimates of statistical uncertainties
due to low-number statistics followed in \citet{mf95}.  Alternative
parametric and non-parametric methods have been also applied to
determine the TRGB magnitude in a number of stellar systems
\citep[see, e.g.][]{sc:1998, cioni, makarov, conn}.

Irrespective of the technique employed to determine the observed TRGB
magnitude, the use of the TRGB as distance indicator has been
traditionally based on $I$-band photometry, in the assumption that the
observed RGB of a given stellar system has an age comparable to the
age of Galactic globular clusters, or in any case larger than 4-5~Gyr
\citep[see, e.g.,][]{dca:1990, lfm:1993, scw02}.  The reason is that
in this age range $M_{bol}^{\rm TRGB}$ is roughly constant at a given
[M/H] and for [M/H] ranging between $\sim -$2.0 and $\sim
-$0.7, $M_{bol}^{\rm TRGB}$ is proportional to $\sim -0.18\,$[M/H],
whilst the bolometric correction to the $I$ band $BC_I$ is
proportional to $\sim -$0.24\,($V-I$) \citep{dca:1990}.  Given that the
($V-I$) colour of the TRGB varies approximately as 0.57\,[M/H], $BC_I$
is then proportional to $\sim -0.14\,$[M/H], hence $M_{I}^{\rm
  TRGB}$=$M_{bol}^{\rm TRGB} - BC_I$ is almost independent of [M/H].
Values of $M_{I}^{\rm TRGB}\sim -4.0$ are generally employed, without or
with a mild dependence on [M/H] \citep[see, e.g., the theoretical
  calibration by][]{sc:1997}.
 
As discussed in \citet{bsh04}, \citet{sg:2005}, \citet{csbook:2013},
applying this type of TRGB calibrations to galaxies with a star
formation that reaches ages below 4-5~Gyr and/or metallicities above
[M/H]$\sim -$0.7, can cause sizable systematic errors on their derived
distance, as recently confirmed empirically by \citet{gpgc:2016}.  To
circumvent this problem it is appropriate to use an
$M_{I}^{\rm TRGB}-(V-I)^{\rm TRGB}$ relation, whereby the TRGB colour
essentially accounts for the star formation history of the observed
population.  A relationship of this type appears to be smooth and
tight, both empirically \citep{rizzi07} and theoretically, as shown
below.

Figure~\ref{isocolrel} displays the theoretical $M_{I}^{\rm
  TRGB}-(V-I)^{\rm TRGB}$ calibration in Johnson-Cousins filters (here
and in the following plots we consider ages lower than $\sim$15~Gyr,
discarding points corresponding to unrealistically old ages) derived
from our \lq{best choice}\rq calculations (Sect.~\ref{s:standard}),
after applying the empirical bolometric corrections by 
\citet{wl11}.  The upper panel displays with different symbols the
TRGB magnitude and colour for [M/H] larger (up to [M/H]$\sim$0.4) and
lower (down to a lower limit [M/H]$\sim -$2.3) than $-$0.65,
respectively.  In general, decreasing age at constant [M/H] shifts
$M_{I}^{\rm TRGB}$ to bluer colours, while increasing [M/H] at fixed
age has the opposite effect.  The overall shape of the $M_{I}^{\rm
  TRGB}-(V-I)^{\rm TRGB}$ relation displays  maximum brightness around
$(V-I)^{\rm TRGB} \sim 1.7$~mag, corresponding to [M/H]$\sim-$1 and ages
between $\sim$8~Gyr and $\sim$12~Gyr.  For $(V-I)^{\rm TRGB}$ below
$\sim$1.1~mag (corresponding to [M/H]=$-$2.3 and ages below 4~Gyr)
$M_{I}^{\rm TRGB}$ increases at almost constant colour, whilst above
$(V-I)^{\rm TRGB}\sim 1.7$ the TRGB magnitude increases with colour
much more slowly.

The lower panel of Fig.~\ref{isocolrel} displays with different
symbols TRGB magnitudes and colours for ages above 4~Gyr with [M/H]$>
-$0.65 and [M/H]$< -$0.65 (this latter being the
\lq{classical}\rq\ TRGB calibration for distance determinations)
respectively.

   \begin{figure}
   \centering
   \includegraphics[width=9cm]{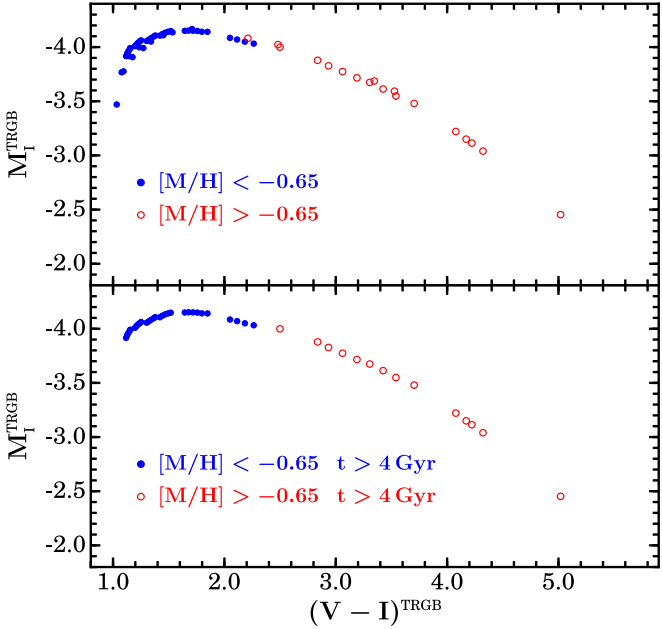}
      \caption{Theoretical $M_{I}^{\rm TRGB}-(V-I)^{\rm TRGB}$
        relation covering the full age and [M/H] range of our
        reference calculations (upper panel), and considering only
        ages above 4~Gyr (lower panel).  Two different [M/H] ranges
        are denoted with different symbols, as labelled.}
         \label{isocolrel}
   \end{figure}

The \lq{classical}\rq\ calibration with our reference models provides
a median $M_{I}^{\rm TRGB}$=$-$4.07 with a total range of
$\pm$0.08~mag around this value. The behaviour of $M_{I}^{TRGB}$ vs
$(V-I)^{\rm TRGB}$ in this age and metallicity range is approximately
quadratic, but assuming a constant average value of $M_{I}^{\rm TRGB}$
is still a decent approximation, as shown by the relatively small
range of values spanned by $M_{I}^{\rm TRGB}$.

It is evident that the $M_{I}^{\rm TRGB}-(V-I)^{\rm TRGB}$
relationship is very smooth, irrespective of any [M/H] and age
selection, and for $(V-I)^{\rm TRGB}$ larger than $\sim$1.1~mag it can
be used as distance indicator of any stellar population that displays
a well populated RGB.

\begin{figure}
   \centering
   \includegraphics[width=9cm]{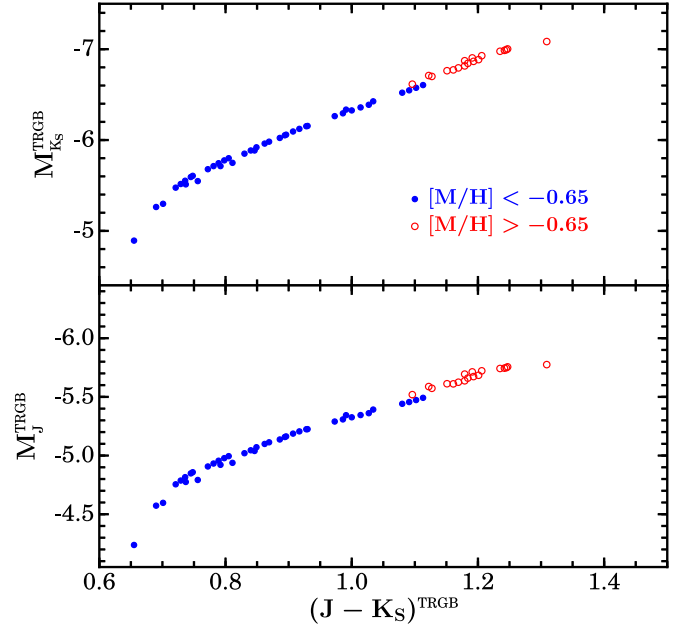}
      \caption{As the upper panel of Fig.~\ref{isocolrel}, but for the
        $M_{K_s}^{\rm TRGB}-(J-K_s)^{\rm TRGB}$ (upper panel) and
        $M_{J}^{\rm TRGB}-(J-K_s)^{\rm TRGB}$ relations (lower panel)
        in the 2MASS photometric system.}
         \label{isocolrelIR}
   \end{figure}

The same holds true in infrared filters. Figure~\ref{isocolrelIR}
displays $M_{K_s}^{\rm TRGB}-(J-K_s)^{\rm TRGB}$ and $M_{J}^{\rm
  TRGB}-(J-K_s)^{\rm TRGB}$ relations in the 2MASS photometric system,
derived from our reference calculations and the bolometric corrections
by \citet{wl11}, after applying the \citet{carpenter} transformations
from the Bessell-Brett system to the 2MASS one.  The magnitude-colour relationships are
again very smooth, with the added bonus of a lower sensitivity to
reddening and a much more linear behaviour compared to the $VI$ bands.
Notice also how the range of the TRGB absolute magnitude in
$J$ is reduced compared to the case of the $K$  band.
\citet{wu14} have also derived empirical tight and smooth
relationships in the corresponding infrared filters of the WFC3-$HST$
system ($F110W$ and $F160W$).

In the following we will consider $VI$ and $JK$ absolute
magnitude-colour relationships as the most reliable way to determine
TRGB distances, and discuss the uncertainties in their theoretical
calibration, stemming from both stellar model and bolometric
correction uncertainties.

\subsection{Theoretical uncertainties on TRGB distances: stellar models}\label{s:theory}

To discuss the effect of varying stellar input physics on theoretical
TRGB distance calibrations, we have calculated additional sets of
models by varying one \lq{ingredient}\rq\ at a time compared to the
reference GARSTEC set based on the physics inputs listed in
\ref{s:physics}, with a solar calibrated mixing length $\alpha_{\rm
  MLT}$ and $\Delta t_{\rm min}$=1~kyr.

More in detail, the additional sets include these changes:

\begin{enumerate}
\item{variation of the $3\alpha$ reaction rate by $\pm$10\%;}
\item{use of NACRE reaction rates instead of those by SFII but the
  LUNA rate for the ${\rm ^{14}N(p, \gamma)^{15}}$ reaction;} 
\item{use of NACRE rates for all reactions} 
\item{\emph{weak} screening instead of the appropriate value;}
\item{variation of the mixing length $\alpha_{\rm MLT}$ by $\pm0.2$.}
\end{enumerate}

\begin{figure}
   \centering
   \includegraphics[width=9cm]{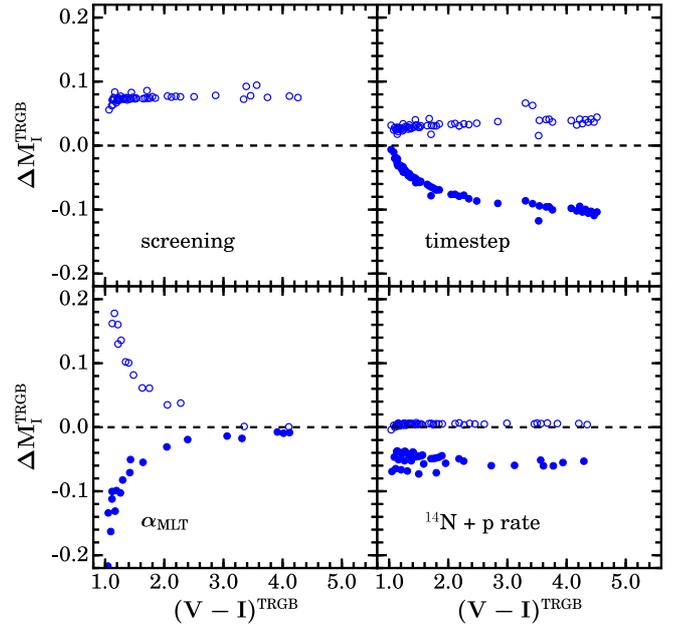}
      \caption{Difference of $M_{I}^{\rm TRGB}$ at fixed $(V-I)^{\rm
        TRGB}$ colour when varying -- one at a time -- the labelled
        physics inputs and computational timestep (see text for
        details). Filled circles in the various panels display the
        cases with $\Delta t_{\rm min}$=3~kyr timesteps, $\Delta
        \alpha_{\rm MLT}=+$0.2, and NACRE reaction rates for all
        nuclear reactions (including the ${\rm ^{14}N(p,
          \gamma)^{15}O}$ reaction), respectively. Open circles display the cases 
         of weak screening, $\Delta t_{\rm min}$=0.3~kyr,  $\Delta
        \alpha_{\rm MLT}=-$0.2, and NACRE rates plus the LUNA rate for the 
        $^{14}$N+p reaction, respectively.}
         \label{vimodels}
   \end{figure}

Figure~\ref{vimodels} compares the corresponding variations of
$M_{I}^{\rm TRGB}$ (the variations of the
$K_s$ and $J$ TRGB magnitudes at fixed $(J-K)$ follow the same
qualitative and quantitative pattern).  The displayed differences are
in the sense {\sl TRGB with varied input $-$ reference GARSTEC TRGB},
and have been determined by interpolating the results for the
reference model set to the colours of each of the calculations with
varied inputs.  We do not show the results for the first case because the
variations of the TRGB brightness are essentially zero. We add instead
the case of models calculated with $\Delta t_{\rm min}$=3~kyr 
(standard GARSTEC calculations) 
and 0.3~kyr, presented in Sect.~\ref{s:stability}.

The choice of the LUNA rate for the ${\rm ^{14}N(p, \gamma)^{15}}$
reaction and NACRE rates for all other reactions has a negligible
effect, at the level of $\sim$0.01~mag on average. In case of NACRE
rates also for the critical ${\rm ^{14}N(p, \gamma)^{15}}$ reaction
the impact is, on average, at the level of 0.05~mag for both $VI$ and
$JK$ calibrations, the TRGB being brighter if the NACRE rate is
employed.

The choice of the timestep and treatment of electron screening have
also an appreciable effect. The use of \emph{weak} screening instead
of the appropriate choice between \emph{strong}, \emph{intermediate}
and \emph{weak} cases, makes the TRGB fainter (because of smaller
He-core masses at the He-flash) by typically around 0.08~mag\footnote{Note that 
\emph{weak}, \emph{intermediate} and \emph{strong} refer to the amount
of electron degeneracy but, in fact, \emph{weak} screening enhances
reaction rates more strongly than intermediate screening in RGB
cores.}. As for 
the timestep, the standard $\Delta t_{\rm min}$=3~kyr
provide TRGB magnitudes systematically brighter (at fixed colour) compared 
to the reference results with $\Delta t_{\rm min}$=1~kyr. There is a
clear strong trend of these differences with colour, in the sense of
larger differences (up to $\sim$0.10~mag) for redder colours
(higher metallicities).  The difference with results for models
calculated with the shortest timesteps is very small, of
the order of just 0.03~mag on average (see also the discussion in
Sect.~\ref{s:stability}).

Finally, the effect of mixing length variations is never
negligible. Given that the bolometric luminosity of the TRGB is
unaffected by the choice of $\alpha_{\rm MLT}$, the differences seen
in Fig.~\ref{vimodels} are due to the change of the TRGB effective
temperature with changing $\alpha_{\rm MLT}$, and the resulting change
of colours and bolometric corrections. 
In practice, variations of $\alpha_{\rm MLT}$ can be understood as a proxy 
for uncertainties in the $\teff$ scale of RGB models. As discussed in \citet{cassisi:2010}, 
these uncertainties are of the order of 200-300~K, corresponding to  
$\Delta \alpha_{\rm MLT} \sim$ 0.20$-$0.30, depending on the metallicity of models. Therefore, 
by considering a range $\Delta  \alpha_{\rm MLT}=\pm 0.2$, we set a rather conservative
estimate of the $\teff$ uncertainty on the absolute magnitude of the TRGB. The decrease of 
$\alpha_{\rm  MLT}$ makes the TRGB fainter in the I and J bands (the opposite is true for an increase
of $\alpha_{\rm MLT}$), the magnitude change increasing steeply with decreasing colour, up to $\sim$0.2~mag. As stated above, this is probably an overestimate of the true uncertainty, which is more likely to be $\pm 0.1$ or smaller (lower left panel in Fig.\,\ref{vimodels}).

These systematic effects are roughly additive, so that the combination 
``$\Delta t_{\rm min}$=0.3~kyr''  + ``weak screening'' + ``\citet{adelb}
$^{14}$N+p rate'' + ``increased $\alpha_{\rm MLT}$'' would provide TRGB
magnitudes on average $\sim 0.25-0.30\, {\rm mag}$ fainter than the
combination  ``$\Delta t_{\rm min}$=3~kyr''  + ``full screening'' + ``NACRE
$^{14}$N+p rate'' + ``decreased $\alpha_{\rm MLT}$'', at the extreme
ends of the magnitude-colour calibration.

Our \lq{best choice}\rq\ set of TRGB models is the one defined in
Sect.~\ref{s:standard}, and provides TRGB magnitudes
intermediate between the brighter and the fainter combinations
described above.


An additional potential source of systematics for theoretical TRGB
magnitude-colour calibrations is the choice of the bolometric
corrections (BCs) applied to the TRGB bolometric magnitudes and
effective temperatures predicted by stellar model calculations.

\subsection{Theoretical uncertainties on TRGB distances: bolometric corrections}\label{s:bolo} 

Whereas in case of the model calculations it is possible, at least in
principle, to identify the best combination of input physics/numerics,
it is extremely difficult to select a priori the \lq{best}\rq\ set of
BCs.  In the following we will consider the empirical BCs by
\citet{wl11} that we have already used in the previous plots, plus the
semiempirical BCs by \citet{westera02} and the theoretical results
from the MARCS \citep{marcs} and PHOENIX \citep{dartm} model
atmospheres\footnote{For the MARCS BCs we employed the routine provided by 
\citet{casa:2014}, whereas PHOENIX BCs have been provided by A. Dotter 
(private communication).}. The \citet{westera02} BCs do not include metallicities
[M/H] below $-$2.0, whilst the MARCS results do not include
surface gravities log($g$)$<$0.0, therefore they do not cover the
oldest solar metallicity TRGBs and the supersolar ones.

\begin{figure}
   \centering
   \includegraphics[width=9cm]{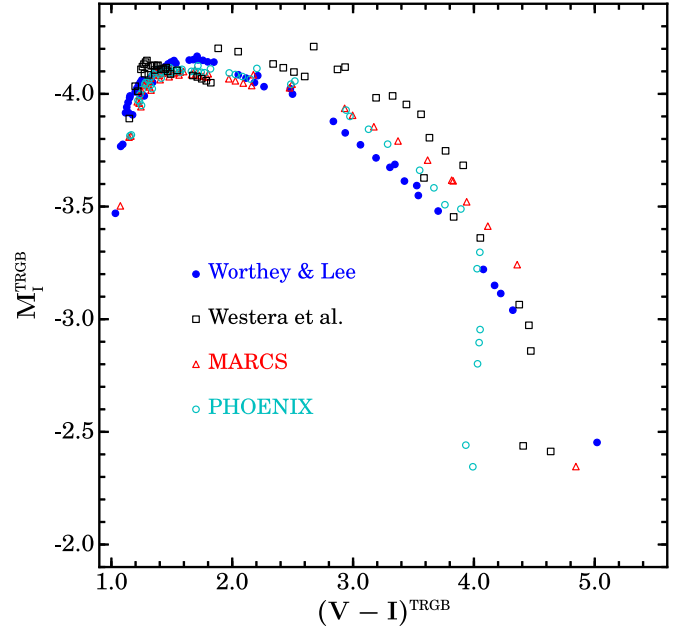}
      \caption{Theoretical $M_{I}^{\rm TRGB}-(V-I)^{\rm TRGB}$
        relation covering the full age and [M/H] range of our
        reference calculations employing the fours labelled sets of BCs}
         \label{vibcs}
   \end{figure}

\begin{figure}
   \centering
   \includegraphics[width=9cm]{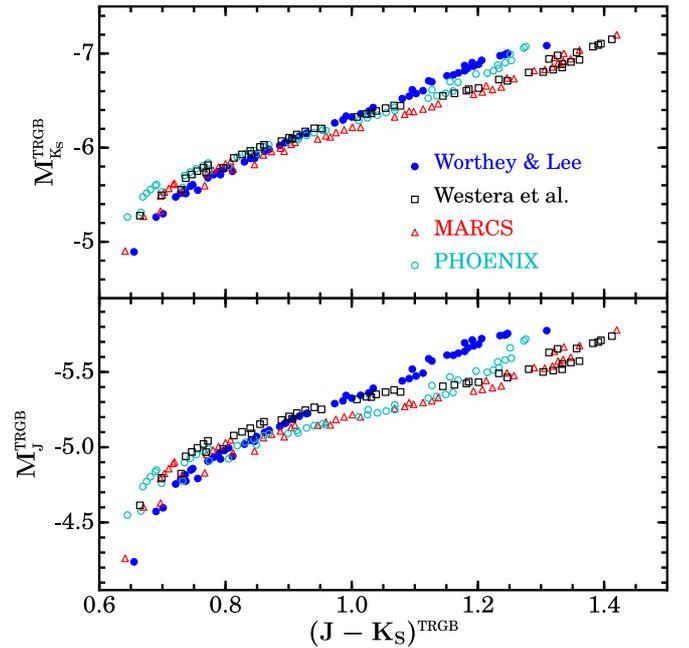}
      \caption{As Fig.~\ref{vibcs}, but for the 
        $M_{K_s}^{\rm TRGB}-(J-K_s)^{\rm TRGB}$ (upper panel) and
        $M_{J}^{\rm TRGB}-(J-K_s)^{\rm TRGB}$ relations (lower panel)
        in the 2MASS photometric system.}
         \label{jkbcs}
   \end{figure}

Figure~\ref{vibcs} and ~\ref{jkbcs} compare the $VI$ and $JK$ absolute
magnitude-colour calibrations obtained with our best choice
theoretical models and these sets of BCs. The differences are not
negligible and become substantial in certain colour ranges.  In case
of the $M_{I}^{\rm TRGB}-(V-I)^{\rm TRGB}$ calibrations, $M_{I}^{\rm
  TRGB}$ differences at fixed $(V-I)^{\rm TRGB}$ increase with
increasing colours. When $(V-I)^{\rm TRGB}$ is between $\sim$2.5 and
$\sim$4.0 ([M/H] approximately between $-$0.6 and solar) the
differences reach $\sim$0.3~mag, whilst at bluer colours they are
typically of the order of 0.1-0.15~mag.  Above $(V-I)^{\rm
  TRGB}\sim$4.0 the slopes of the magnitude-colour relationships are
radically different.  The \citet{westera02} BC produce sizable
discontinuities in the magnitude-colour relationship due to the
stronger dependence of the BC to the $I$-band on metallicity,
compared to the other sets of BCs.

A word of caution is in place regarding the use of $VI$. The point 
where He-core ignition occurs always corresponds to the point with the
largest bolometric luminosity, but this does not necessarily coincide 
with the brightest point, i.e. the TRGB, in the $I$ band, because of the dependence 
of the BC on effective temperature. In fact, 
the He-core ignition point and the TRGB are different with the Westera 
transformation for $(V-I) > 3.0$ (the point of He-core ignition is
fainter than the TRGB) while, for the Worthey \& Lee transformation 
the He-flash always corresponds to the TRGB. For the MARCS models, 
the brightest TRGB in the $I$ band is reached around $\teff \sim 3400$~K, 
with the TRGB becoming dimmer in the $I$ band at lower $\teff$. Therefore, 
depending on the set of BCs that are used, the He-flash can be fainter than the TRGB and hence, 
from a theoretical point of view, the TRGB is much less well defined when 
using $VI$ photometry. Observationally, the $M_{I}-(V-I)$ diagram of 47~Tuc in \citet{bell04}
shows an almost constant $M_{I}$ at the reddest $V-I$ colors and indicates that this 
caveat is relevant for old populations more metal rich than $\feh \approx -0.5$.

In the infrared filters the various sets of BCs produce different
overall slopes of the magnitude-colour relationships.  Differences of
$M_{J}^{\rm TRGB}$ or $M_{K}^{\rm TRGB}$ at fixed $(J-K)^{\rm TRGB}$
increase towards larger colours (higher metallicities), reaching
$\sim$0.30~mag in $J$ and $K$ at $(J-K_s)^{\rm TRGB} > 1.0-1.1$ ([M/H]
above $\sim -$0.65).  The closest agreement is in the colour range
0.8$<(J-K_s)<$1.0, where differences are at most of the order of
$\sim$0.1~mag. Below $(J-K_s)^{\rm TRGB}\sim$0.8 (corresponding to
[M/H]$\sim-$1.5 to $-$1.8, depending on age) the effect
of the choice of BCs increases again, up to $\sim$0.2~mag.
In $JK$ photometry, the He-flash is always the brightest point, i.e. it 
always coincides with the TRGB at all $(J-K)$ colors for these sets of BCs.

\begin{figure}
   \centering
   \includegraphics[width=9cm]{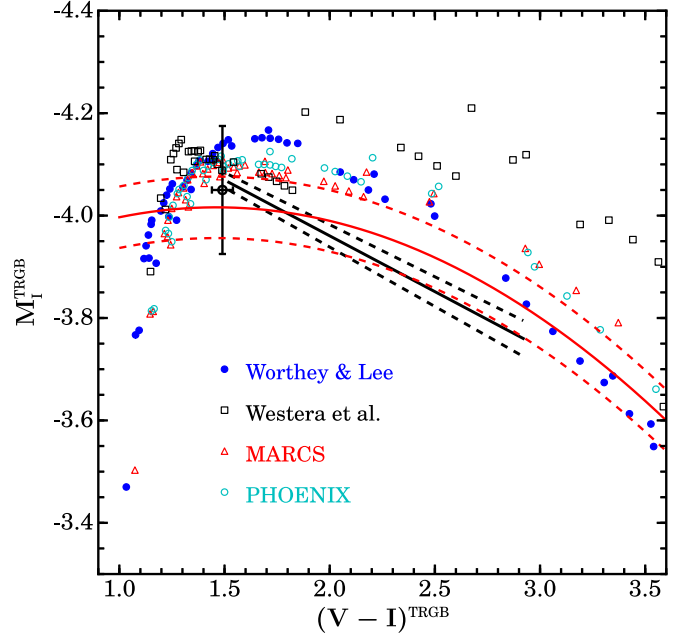}
      \caption{As Fig.~\ref{vibcs}, but  
        including the empirical value obtained for the Galactic globular cluster 
       $\omega$~Cen (open circle with error bars), and the  
       empirical relationships by \cite{rizzi07} and \cite{jang:17} 
       based on samples of local galaxies (black and red lines,
       respectively, see text for details).  
       The solid lines denote mean values of the relationships, while the dashed lines 
       denote, respectively, the brighter and lower limits according to the 1$\sigma$ 
       uncertainty on the slope and zero point of \cite{rizzi07}
       results, and the 1$\sigma$ uncertainty on the zero point of
       \cite{jang:17} calibration.} 
         \label{vibcstest}
   \end{figure}

Existing empirical constraints on the TRGB brightness do not help
discriminating amongst the various sets of BCs in the $VI$ diagram,
and actually highlight inconsistencies, as shown by
Fig.~\ref{vibcstest} that displays the $M_{I}^{\rm TRGB}-(V-I)^{\rm
  TRGB}$ calibrations of Fig.~\ref{vibcs} together with the empirical
estimate for the Galactic globular cluster $\omega$~Cen \citep{bell01,
  bell04} based on the cluster eclipsing binary distance.  The error
bar on $M_{I}^{\rm TRGB}$ is still substantial and cannot put strong
constraint on the theoretical calibration.  We do not include an
analogous estimate for the more metal-rich Galactic globular 47~Tuc,
because of large error bars and the need to apply theoretical
corrections to infer the real TRGB location from the observed one, due
to the paucity of bright RGB stars that bias the TRGB detection
\citep[see][for details]{bell04}.

In the same figure we also display the empirical \citet{rizzi07}
relationship, that has been applied by \citet{jrt:2009} to determine
distances to a large sample of galaxies in the Local Volume, and the
recent \cite{jang:17} empirical calibration.  The slope of the
\citet{rizzi07} relationship was estimated by determining the apparent
TRGB magnitude vs colour relation in a sample of galaxies, and
averaging the slopes. The zero point determination was based on
horizontal branch distances to IC1613, NGC185, the Sculptor and Fornax
dwarf spheroidal galaxies, and M33 based on the \citet{carr00} HB
absolute magnitude vs [M/H] relationship. This relation was obtained
from fitting main sequence distances to a sample of Galactic globular
clusters using subdwarfs with Hipparcos parallaxes. The crucial
difficulty here is to assign a [M/H] to the HB stars in the
galaxy. The authors employed a $(V-I)^{\rm TRGB}$-[M/H] relationship
based on Galactic globulars \citep[see][for details]{rizzi07} and
assigned to the observed HB the [M/H] derived from the measured
$(V-I)^{\rm TRGB}$ colour. 

The quadratic form of the \cite{jang:17} calibration is also based on the
apparent TRGB magnitude vs colour relations in a galaxy sample
\citep[different from][]{rizzi07}, with the zero point set by
averaging the zero points obtained from the LMC (distance from
eclipsing binaries) and NGC4258 maser distance (the two zero points
are however almost identical, see the paper for details).  These two
empirical calibrations are clearly different from the theoretical
predictions, but also mutually different, especially in terms of the
trend of the TRGB magnitude with colour.

\citet{rizzi07} TRGB magnitudes are generally fainter than the models,
with a steeper slope compared to our results, irrespective of the
adopted BCs.  At $(V-I)^{\rm TRGB}2.0 - 2.5$ differences in
$M_{I}^{\rm TRGB}$ can reach up to 0.2-0.3~mag, depending on the
chosen set of BCs.  The calibration by \cite{jang:17} produces usually
fainter TRGBs compared to the models (apart from the bluest colours),
but the trend with colour, at least for $(V-I)^{\rm TRGB}>1.5$, is
consistent with the theoretical calibration employing MARCS BCs.

\begin{figure}
   \centering
   \includegraphics[width=9cm]{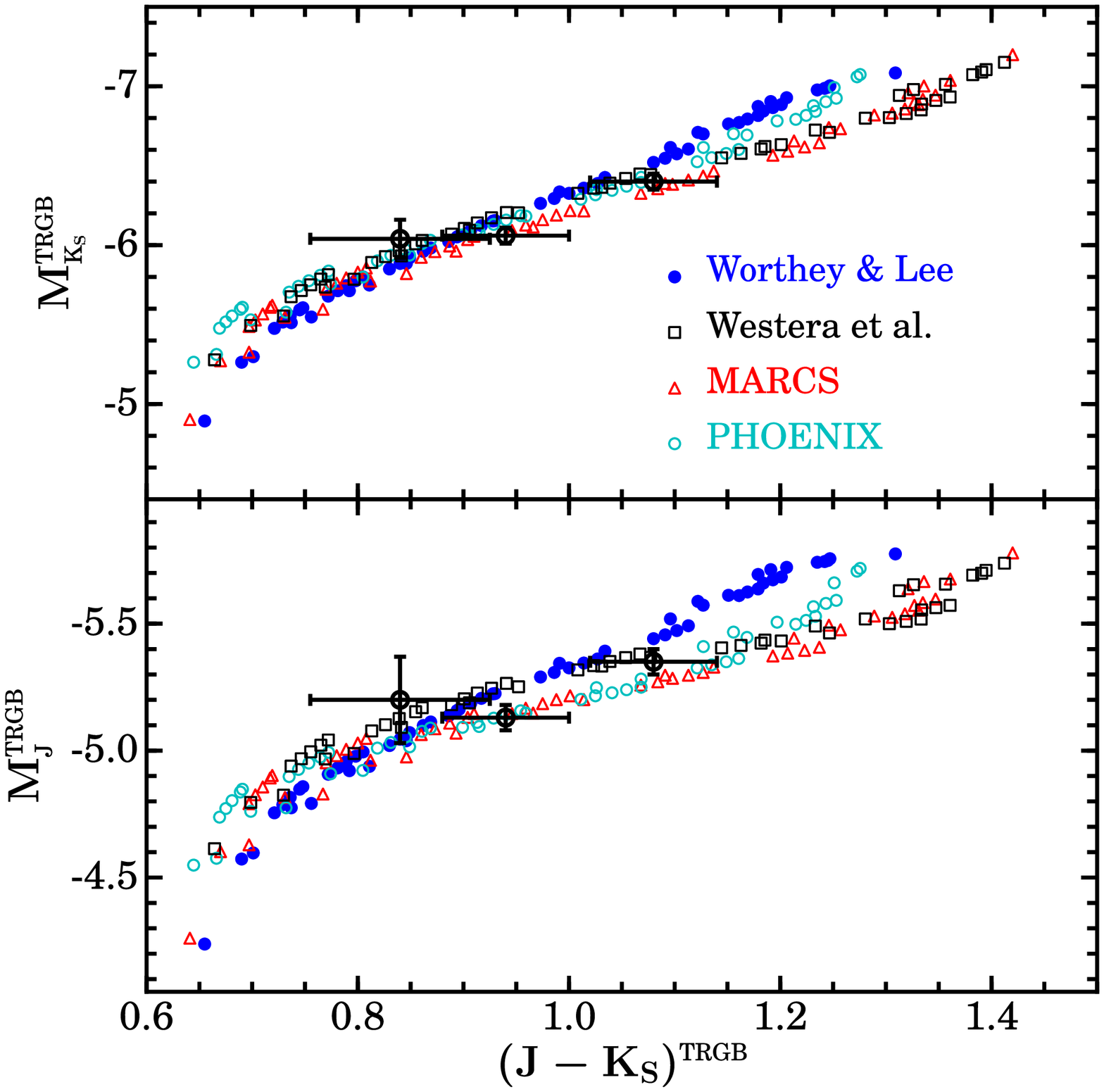}
      \caption{As Fig.~\ref{jkbcs}, but including the empirical values obtained 
      for $\omega$~Cen, the SMC and LMC.}
         \label{jkbcstest}
   \end{figure}

For the infrared the situation is different. 
Figure~\ref{jkbcstest} shows the $JK$ absolute magnitude-colour calibrations of 
Fig.~\ref{jkbcs} together with the empirical estimate for the Galactic globular 
cluster $\omega$~Cen, and the mean values for the LMC and SMC obtained 
from a number of fields within each galaxy. These latter two 
estimates \citep[from][]{gpgc:2016} rely on the eclipsing binary distances to 
the Magellanic Clouds. The three constraints fall in the colour range where 
the differences due to the choice of BCs are minimized, 
and within the errors are consistent with all four sets of BCs.

   \begin{figure}
   \centering
   \includegraphics[width=9cm]{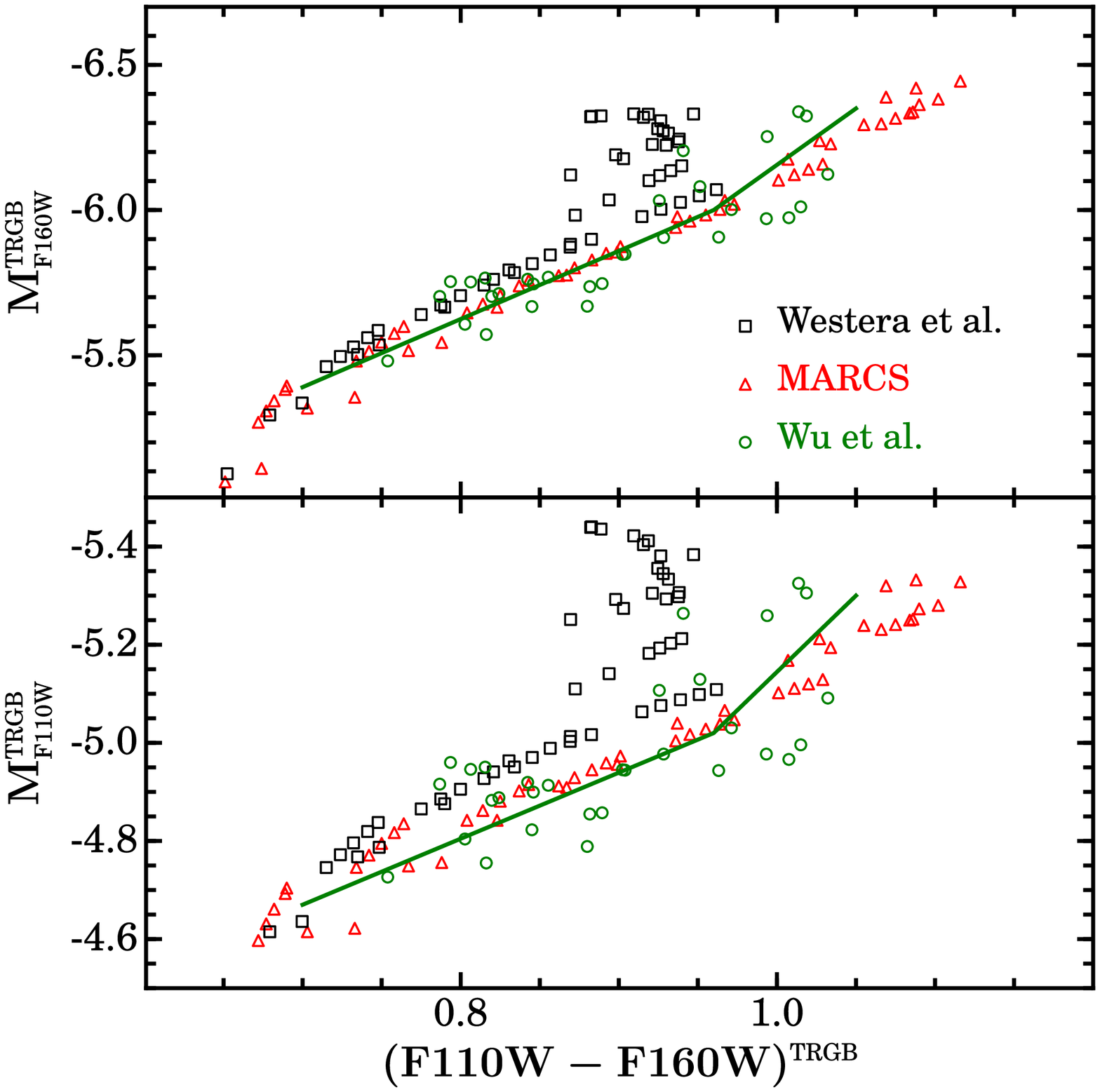}
      \caption{As Fig.~\ref{jkbcs}, but for the WFC3-$HST$ filters $F110W$ and $F160W$. 
      We include the theoretical 
       calibrations based on \citet{westera02} and MARCS BCs, and the empirical one 
       (analytical formula plus the calibrating datapoints without
       error bars) by \citet{wu14}.} 
         \label{hstbcstest}
   \end{figure}

Finally, Fig.~\ref{hstbcstest} shows the magnitude-colour calibrations
in the WFC3-$HST$ filters $F110W$ and $F160W$, similar to $J$ and $K$,
compared to the empirical calibration by \citet{wu14} based on a
sample of local galaxies, and fully consistent (in zero point) with
the \citet{rizzi07} calibration in $VI$.  We plot only theoretical
results employing \citet{westera02} and MARCS BCs, for we do not have
PHOENIX and \citet{wl11} BCs available in these filters.  The
theoretical results employing MARCS BCs are in good agreement with the
empirical calibration. Some discrepancies appear at $(F110W-F160W)$
above $\sim$0.95, where however the empirical datapoints show a large
dispersion.  \citet{wu14} determined 1$\sigma$ dispersions around the
mean relationship by 0.05 and 0.07~mag in $F110W$ and $F160W$
respectively, when $(F110W-F160W)$ is below 0.95. For larger colours
the dispersions are equal to 0.12 and 0.09~mag in $F110W$ and $F160W$
respectively.

\begin{table*}
\caption{Analytic fits to our reference calibrations of the RGB-tip magnitude-color relations. \label{tab:fits}}
\centering
\begin{tabular}{lll}
\hline \\
$M_I^{\rm TRGB}$ & $ = \left\{ \begin{array}{l} 
 -4.090 + 0.086\,[(V-I)^{\rm TRGB}-1.4] + 4.721\,[(V-I)^{\rm TRGB}-1.4]^2 \\ 
 -4.090 + 0.017\,[(V-I)^{\rm TRGB}-1.4] + 0.036\,[(V-I)^{\rm TRGB}-1.4]^2 \\ 
 -4.037 + 0.087\,[(V-I)^{\rm TRGB}-2.4] + 0.158\,[(V-I)^{\rm TRGB}-2.4]^2 \\ 
 \end{array} \right. $ & 
$ 
\begin{array}{l}
 1.00 < (V-I)^{\rm TRGB} < 1.40  \\
 1.40 \leq (V-I)^{\rm TRGB} < 2.40  \\
 2.40 \leq (V-I)^{\rm TRGB} < 4.50  \\
\end{array}$
  \\ \\
$M_{K_S}^{\rm TRGB}$ & $ = \left\{ \begin{array}{l}
 -5.722 - 2.386\,[(J-K)^{\rm TRGB}-0.76] + 34.694\,[(J-K)^{\rm TRGB}-0.76]^2 \\ 
 -5.722 - 1.811\,[(J-K)^{\rm TRGB}-0.76] -  0.517\,[(J-K)^{\rm TRGB}-0.76]^2  \\
\end{array} \right. $ & 
$ 
\begin{array}{l}
 0.60 < (J-K)^{\rm TRGB} < 0.76  \\
 0.76 \leq (J-K)^{\rm TRGB} < 1.50  \\
\end{array}$
\\ \\
$M_{J}^{\rm TRGB} $ & $ = \left\{ \begin{array}{l}
 -4.962 - 1.386\,[(J-K)^{\rm TRGB}-0.76] + 34.694\,[(J-K)^{\rm TRGB}-0.76]^2 \\ 
 -4.962 - 0.811\,[(J-K)^{\rm TRGB}-0.76] - 0.517\,[(J-K)^{\rm TRGB}-0.76]^2  \\
\end{array} \right. $ & 
$ 
\begin{array}{l}
 0.60 < (J-K)^{\rm TRGB} < 0.76  \\
 0.76 \leq (J-K)^{\rm TRGB} < 1.50  \\
\end{array}$
 \\ \\
$F110W^{\rm TRGB} $ & $ =  \left\{ \begin{array}{l}
 -4.630 - 9.525\,[(F110W-F160W)^{\rm TRGB}-0.68]  \\ 
 -4.630 - 1.511\,[(F110W-F160W)^{\rm TRGB}-0.68] \\ 
\end{array} \right. $  & 
$ 
\begin{array}{l}
 0.60 < (F110W-F160W)^{\rm TRGB} < 0.68  \\
 0.68 \leq (F110W-F160W)^{\rm TRGB} < 1.20  \\
\end{array}$
 \\ \\
$F160W^{\rm TRGB} $ & $ =  \left\{ \begin{array}{l}
 -5.310 - 10.525\,[(F110W-F160W)^{\rm TRGB}-0.68]  \\ 
 -5.310 - 2.511\,[(F110W-F160W)^{\rm TRGB}-0.68] \\ 
\end{array} \right. $  & 
$ 
\begin{array}{l}
 0.60 < (F110W-F160W)^{\rm TRGB} < 0.68  \\
 0.68 \leq (F110W-F160W)^{\rm TRGB} < 1.20  \\
\end{array}$
\\
\hline 
\end{tabular}
\tablefoot{Calibrations are based on \texttt{MARCS} colors and BCs and are valid for  $10^{-4} \leq Z \leq 
0.04$ except for the $M_I^{\rm TRGB}$ vs $(V-I)^{\rm TRGB}$ relation, valid up to $Z \approx 0.02$. 
Physical inputs in the models, including initial composition, are described in Sect.~\ref{s:physics}. 
Models do not include $\alpha$-enhancement (see text for discussion). }
\end{table*}

Considering together the comparisons of Figs.~\ref{jkbcstest} and
\ref{hstbcstest}, the MARCS BCs seem to be well suited to be used for
TRGB modelling in the infrared (up to metallicities around solar).
The situation in the $VI$ diagram is less clear; the \citet{rizzi07}
results (that should be consistent with the IR empirical calibration
discussed above) cannot be reproduced with any set of BCs, whilst the
TRGB of $\omega$~Cen does not provide an additional strong constraint
due to the relatively large associated error bar.  On the other hand,
the alternative empirical calibration by \cite{jang:17} displays a
trend with colour, for $(V-I)^{\rm TRGB}>$1.5, generally consistent
with our models complemented by MARCS BCs, and a zero-point offset
only very slightly larger than the 1$\sigma$ uncertainty (0.06~mag)
associated to \cite{jang:17} TRGB magnitudes. 

\subsection{Reference calibrations} \label{sec:reference}

Based on discussions in previous sections, we present reference calibrations for the RGB-tip in 
different color-magnitude combinations. These are based on stellar models
computed with the concordance input physics described in Sect.\,\ref{s:physics}, short 
minimum timestep (1~kyr; Sect.\,\ref{s:stability}) and the MARCS BCs. Calibrations are presented in Table~\ref{tab:fits} as a set of analytic fits to facilitate their use. The dispersion between models and the fits are 0.02 for $M_I^{\rm TRGB}-(V-I)^{\rm TRGB}$ and 0.04 for the other relations. This is, in all cases, much smaller than the 0.12 uncertainty in the theoretical RGB-tip magnitudes discussed in Sect.\,\ref{s:theory}.

For reference, Table~\ref{tab:models} includes some properties of stellar models at the TRGB, including age, surface helium abundance ($Y_{\rm S}$), luminosity, $\teff$ and $M_{\rm V}^{\rm TRGB}$. 

\begin{table*}
\caption{Summary of TRGB properties for selected models. \label{tab:models}}
\centering
\begin{tabular}{ccccccccc}
\hline \\
$Z$ & $Y$ & Mass & Age & $M_{\rm c}^{\rm He}$ & $Y_{\rm S}$ & $\log{\ltip}$ & $\log{T_{\rm eff}^{\rm TRGB}}$ &
$M_{\rm V}^{\rm TRGB}$\\
\hline 
\rule{0pt}{2.5ex} 0.0001 &  0.245 & 0.80 & 12.72 & 0.500 & 0.253 & 3.250 & 3.651 & -2.764 \\
'' & '' & 0.90 & \ 8.46 & 0.498 & 0.257 & 3.237 & 3.654 & -2.765 \\
'' & '' & 1.00 & \ 5.92 & 0.494 & 0.260 & 3.217 & 3.658 & -2.751 \\
\hline
\rule{0pt}{2.5ex} 0.0010 & 0.246 & 0.80 & 13.82 & 0.419 & 0.259 & 3.336 & 3.598 & -2.502 \\
'' & '' & 0.90 & \ 9.13 & 0.488 & 0.263 & 3.327 & 3.603 & -2.551 \\
'' & '' & 1.00 & \ 6.34 & 0.485 & 0.266 & 3.316 & 3.608 & -2.595 \\
\hline
\rule{0pt}{2.5ex} 0.0040 & 0.251 & 0.85 & 13.76 & 0.485 & 0.269 & 3.390 & 3.552 & -1.876 \\
'' & '' & 0.95 & 9.24 & 0.483 & 0.272 & 3.385 & 3.558 & -2.032 \\
'' & '' & 1.00 & 7.70 & 0.482 & 0.273 & 3.382 & 3.561 & -2.095 \\
\hline 
\end{tabular}
\end{table*}

\section{Discussion} \label{s:discussion}

The maximum brightness of low-mass stars along the RGB is a
reference point in any CMD. It can be used for
various purposes, ranging from absolute and relative distance
determinations to non-standard cooling processes of degenerate cores,
but also for age determinations of old populations, where its weak
sensitivity to age can be used in determining distances and thus allow 
using other time-dependent CMD features to date it. To
achieve a high accuracy in the results, one clearly needs an absolute
calibration of the TRGB luminosity and its --even if minor-- dependence on age and
metallicity, both in terms of bolometric flux and magnitudes in photometric passbands. 
The present work is concerned with the theoretical
prediction for the TRGB brightness, which depends
both on the physics employed in the theoretical models, and on the
numerical and technical details of stellar evolution codes. We have,
for the first time, investigated a wide range of stellar masses (from
0.8 to 1.4~$M_\odot$) and metallicities (from $1\times 10^{-4}$ to
$4\times 10^{-2}$). This implies an age range from about 2 to
16~Gyr. 

We started with comparing results for models computed with nominally
very similar standard physics by two standard evolutionary codes,
\texttt{GARSTEC} \citep{weiss:2008} and the \texttt{BaSTI} code
\citep{pietrinferni:2004}. The results, presented in
Sect.~\ref{s:initcomp}, are representative of the variation in
theoretical predictions when using different model resources. They
actually may represent a lower limit to the systematic uncertainties
one should take into account when comparing models to observations. We
showed that the differences can be appreciable for the lowest
metallicities and higher masses, and larger than expected from the so 
far investigated model dependencies, discussed in
Sect.~\ref{s:review}, which were almost exclusively done for older
stars of masses below $1\, M_\odot$.

We then (Sect.~\ref{s:physics}) adjusted the physics 
between the two codes, identifying the main sources for the
differences: these were the ${\rm ^{14}N(p, \gamma)^{15}O}$ CNO-cycle
bottleneck reaction and the electron screening of the
$3\alpha$-rate. With these adjustments, the differences shrunk by up
to a factor of 4 for the low-metallicity, high-mass cases, but in fact
increased for the highest metallicity.

In the next step numerical details were scrutinized
(Sect.~\ref{s:stability} and the temporal
resolution on the upper RGB identified as the --almost sole-- 
remaining issue. We determined the maximum timestep allowed for
converged resolution to be a few hundred years. This resulted in an
overall excellent agreement between the two model sets of order
0.1~dex in ${\rm log} (L_{\rm TRGB}/{\rm L_\odot})$ except, again, for the higher
masses at the lowest metallicity (Fig.~\ref{fig:intercomp2}). This
difference we could not resolve, but it may be related to the different
evolution of very small convective cores on the main-sequence, the
only obvious difference we found for these models, and may be related
to the detailed treatment of convective boundaries \citep[see
  also][]{gnmm:2014}. Fortunately, such stars are, at least in the Milky
Way, not relevant because of their combination of young age and very
low metallicity. Nevertheless, this issue requires further
investigations, beyond the goal of the present paper.

With the physics and numerics that resulted in an overall excellent
agreement of the TRGB luminosities derived from both codes, we computed and
present in Sect.~\ref{s:standard} a set of models, which we recommend
as a reference for further applications of the TRGB in
astrophysics. We have used canonical, but nevertheless up-to-date
physics, ignoring mass-loss, diffusion, extra-mixing on the RGB, and
rotation. However, all these potentially interesting effects have a
much larger theoretical uncertainty than, for example, the equation of
state or electron conduction opacities, and therefore constitute a
much larger degree of freedom for modellers and their output.

In the second part of the paper (Sect.~\ref{s:distance}) we have 
transposed results for the TRGB $L$ and $T_{\rm Teff}$ into 
magnitudes and colours, in particular 
red and infrared magnitudes, as they are known to be best suited for
employing the TRGB as distance indicator.  
For our reference set of models we show the global
relations between TRGB brightness and colour in $VI$ and $JK_s$ diagrams 
(Figs.~\ref{isocolrel} and \ref{isocolrelIR}, respectively) based 
on one particular theoretical set of bolometric corrections \citep{wl11}.
Comparisons of results obtained with several BC sets show that 
the uncertainties arising from BC calculations  
(Sect.~\ref{s:bolo}) in fact exceed those from the
stellar models. In infrared colours, the use of MARCS BCs gives results that seem 
to be in best agreement with an existing empirical relation, although the
empirical database is not yet stringent enough for a final
conclusion.
In the $VI$ diagram the existing empirical relations are not in mutual agreement, 
and it is very hard to assess which set of BCs, if any, is best suited for 
a theoretical calibration.

The comparison of our theoretical predictions with the recent work by
\citet{vcsrrvw:2013a}, to which we refered in the discussion of known
dependencies (Sect.~\ref{s:review}), shows some noteworthy
differences. They find a theoretical $M_I=-3.99$ for stellar models
appropriate for the globular cluster M5.  Using matching models from
our reference set, we find $M_I = -4.14$ with the \citet{wl11}
transformations, the ones also used by \citet{vcsrrvw:2013a}. This
difference translates into $\deltal \approx 0.06$~dex, with our models
being more luminous. This is confirmed by comparing our results with
their Fig.\,6.  We trace part of the difference, 0.035~dex, to the
treatment of electron screening, for which \citet{vcsrrvw:2013a} use
Salpeter's formulation of weak screening whereas our reference set of
models includes the intermediate screening regime, appropriate for RGB
stars. We emphasize again that screening in the weak degeneracy limit
is in fact stronger than in the intermediate limit
(Sect.\,\ref{sec:screening}), leading to a dimmer RGB-tip.  Much of
the physics used in our and \citet{vcsrrvw:2013a} models is the same
so it is not obvious to identify the cause for the remainder 0.025~dex
difference. But it is possible this might be due to numerical aspects
of the calculations that, as we have found in this work, can have a
strong impact on the theoretical $\ltip$ predictions. Our results
strongly alleviate the tension between predictions of standard stellar
models for the RGB-tip brightness and observations of M5 found by
\citet{vcsrrvw:2013a}. We predict $M_I= -4.14 \pm 0.12$ or $-4.07\pm
0.12$ for the RGB-tip depending on whether the \citet{wl11} or MARCS
BCs are used, well within the 1$\sigma$ observational band $M_I=
-4.17\pm 0.13$ determined by \citet{vcsrrvw:2013a} for M5 that is dominated
by a 0.11 uncertainty in the distance modulus. This would
basically remove any need for additional cooling, and therefore put
very low limits on a possible magnetic dipole moment of neutrinos or
the mass of axions, as investigated by \citet{vcsrrvw:2013a} and
\citet{vcsrrvw:2013b}.

The cosmological significance of a careful calibration of the TRGB
brightness is exemplified by the following: Recently,
\citet{jang:2017b} re-calibrated the Hubble constant $H_0$ using the
TRGB brightness from their own empirical calibration for deriving the
absolute brightness of supernovae of type Ia. They arrived at a value
of $71.17\pm2.03\,({\rm rand.}) \pm 1.94 ({\rm syst.})\, {\rm km}\,
{\rm s}^{-1} \, {\rm Mpc}^{-1}$, slightly lower than previous,
Cepheid-based determinations, but still in conflict with the value
obtained from the most recent CMB analysis, \citep[$H_0 = 66.93 \,
  {\rm km}\, {\rm s}^{-1} \, {\rm Mpc}^{-1}$]{PlanckColl:2016}. Since
their calibration is, depending on colour transformation, fainter than
our theoretical prediction by 0.05-0.10~mag (see
Fig.~\ref{vibcstest}), a change to our calibration would increase the
distance scale and decrease $H_0$ by 1-2\% to a value between $\approx
68 \cdots 70 \, {\rm km}\, {\rm s}^{-1} \, {\rm Mpc}^{-1}$.

We conclude that it is possible, by a careful investigation into the
details of numerical stellar model computations, to produce a
reference set of accurate predictions for the TRGB luminosity. The
major uncertainty in applying this to CMDs, both of simple or
composite stellar populations, appears to lie in the BC scale. 
Much work is needed here to achieve the same
confidence as in the underlying stellar models.

\begin{acknowledgements}
The authors thank the anonymous referee for the effort in preparing an extensive, knowledgeable and constructive report that has contributed to improve this work. We also want to thank Nicol\'{a}s Viaux for providing original data and
for helpful discussions. 
AS is partially supported by ESP2015-66134-R (MINECO). SC acknowledges
partial financial support from PRIN-INAF2014 (PI: S. Cassisi) and 
from grant AYA2016-77237-C3-1-P from the Spanish Ministry of Economy
and Competitiveness (MINECO).
AW and MS are grateful for the hospitality of the Lorentz Centre in
Leiden, where the initial discussions about this subject took place
during one of their workshops.
This research was supported by the Munich Institute for Astro- and
Particle Physics (MIAPP) of  
the DFG cluster of excellence ``Origin and Structure of the Universe''. 
\end{acknowledgements}

\bibliographystyle{aa}
\bibliography{TRGB}

\end{document}